\documentclass[]{aastex631}

\usepackage{graphicx}
\usepackage{multirow}
\usepackage{longtable}
\usepackage{epstopdf}
\usepackage{ulem}
\usepackage{bm}
\usepackage{subfigure}
\usepackage{array}
\usepackage{color}
\usepackage{graphicx}
\usepackage{txfonts}
\usepackage{txfonts}
\usepackage{hyperref}

\newcommand{\PKHY}[1]{{{#1}}}

\shortauthors{He et al.}
\graphicspath{{./}{figures/}}

\begin{document}

\title{Spatially decomposed $\gamma$-ray features surrounding SNR Kes 79 \& PSR J1853+0056}

\author{Xinbo He}
\affiliation{School of Physics and Astronomy, Sun Yat-Sen University, Guangzhou, 510275, China}
\affiliation{CSST Science Center for the Guangdong-Hongkong-Macau Greater Bay Area, Sun Yat-Sen University, Guangzhou, China}

\author{Yudong Cui}
\affiliation{School of Physics and Astronomy, Sun Yat-Sen University, Guangzhou, 510275, China}
\affiliation{CSST Science Center for the Guangdong-Hongkong-Macau Greater Bay Area, Sun Yat-Sen University, Guangzhou, China}

\author{Paul K. H. Yeung}
\affiliation{Institute of Experimental Physics, Department of Physics, University of Hamburg, Luruper Chaussee 149, D-22761 Hamburg, Germany}
\affiliation{Nicolaus Copernicus Astronomical Center, Polish Academy of Sciences, Rabia\'nska 8, 87-100, Toru\'n, Poland}
\affiliation{School of Physics \& Astronomy, University of Southampton, Highfield, Southampton SO17 1BJ, UK}

\author{P. H. Thomas Tam}
\affiliation{School of Physics and Astronomy, Sun Yat-Sen University, Guangzhou, 510275, China}
\affiliation{CSST Science Center for the Guangdong-Hongkong-Macau Greater Bay Area, Sun Yat-Sen University, Guangzhou, China}

\author{Yong Zhang}
\affiliation{School of Physics and Astronomy, Sun Yat-Sen University, Guangzhou, 510275, China}
\affiliation{CSST Science Center for the Guangdong-Hongkong-Macau Greater Bay Area, Sun Yat-Sen University, Guangzhou, China}

\author{Yang Chen}
\affiliation{Department of Astronomy, Nanjing University, Nanjing, 10023, China}
\affiliation{Key Laboratory of Modern Astronomy and Astrophysics, Nanjing University, Ministry of Education, China}

\email{K.H.Yeung@soton.ac.uk,cuiyd@mail.sysu.edu.cn,tanbxuan@mail.sysu.edu.cn}



\begin{abstract}

There have been substantial improvements on Fermi Large Area Telescope (LAT)  data and analysis tools since the last analysis on the mid-aged supernova remnant (SNR) Kes 79 \citep{Auchettl2014}. Recent multi-wavelength studies confirmed its interaction with molecular clouds. About $0.36\degr$ north from Kes 79, a powerful pulsar -- PSR J1853+0056 also deserves our attention.  In this work, we analyse the 11.5-year Fermi-LAT data to investigate the $\gamma$-ray feature in/around this complex region. Our result shows a more significant detection ($\sim$34.8$\sigma$ in 0.1--50~GeV) for this region. With $\ge$5~GeV data, we detect two extended sources --  Src-N (the  brighter one; radius $\approx0.31\degr$) concentrated at the north of the SNR while enclosing PSR J1853+0056, and Src-S (radius $\approx0.58\degr$) concentrated at the south of the SNR.  Their spectra have distinct peak energies ($\sim$1.0~GeV for Src-N and $\lesssim$0.5~GeV for Src-S), suggesting different origins for them.  In our hadronic model that includes the leaked cosmic-rays (CRs) from the shock-cloud collision, even with extreme values of parameters,   SNR Kes 79  can by no means provide enough CRs reaching clouds at Src-N to explain the local GeV spectrum. We propose that the Src-N emission could be predominantly reproduced by a putative pulsar wind nebula (PWN) powered by PSR J1853+0056. On the other hand, our same hadronic model can  reproduce a majority of the  GeV  emission at   Src-S with typical values of parameters, while the three known pulsars   inside Src-S release a total power that is too low to account for half of its $\gamma$-ray emission.

\end{abstract}



\section{Introduction}
\label{sec1}

The supernova remnant (SNR), Kes 79 (a.k.a. G33.6+0.1), is first discovered by Molonglo at 408 MHz and Parkes 64m at 5 GHz \citep{Caswell1975} with a bright central region. The radio morphology of Kes 79 is characterized by multiple distinct outer shells \citep{Velusamy1991, Rho1998}. The X-ray emissions from Kes 79 are observed by Einstein \citep{SeaquistGilmore1982} and ROSAT \citep{SewardVelusamy1995}. The later X-ray observations with Chandra reveal complex spatial structures, which demonstrate spatial coincidence with the radio-continuum shell \citep{Sun2004}. Detailed XMM-Newton data analysis by \citet{Zhou2016} discovered that the diffuse X-ray emissions in the SNR consist of a cold and a hot components, which lead to an estimated SNR age of 4.4-6.7 kyr. Suzaku results by \citet{Sato2016} suggested a two-temperature model as well.

\citet{Zhou2016} also found several bright X-ray filaments at the edge of the SNR, which suggest a scenario of SNR interacting with some molecular clouds (MCs) at 105~km/s (7.1~kpc). Other evidences for the SNR-MC interaction include:  detection of  OH absorption at both  1665 and 1667 MHz \citep[][]{Stanimirovic2003},  detection of a 95 GHz methanol maser \citep[][]{Zubrin2008},  detection of broadened CO lines \citep[][]{Zhou2016,Kilpatrick2016}, and high velocity ejecta fragment found in X-ray \citep{Zhou2016}. 
 
Hadronic $\gamma$-ray sources of SNR-MC interactions can serve as  stopwatches for the escape of cosmic-rays (CRs) from SNRs, which gradually develops from highest-energy particles to lowest-energy particles with time \citep[see][]{Ptuskin2003,Ptuskin2005}. To be specific, for hadronic interactions involving a mid-aged ($>$3~kyr) SNR like Kes 79, we would expect a relatively softer $\gamma$-ray spectrum peaked at a relatively lower energy \citep[see][]{Suzuki2018,Suzuki2020a,Suzuki2020b}. 

Inside SNR Kes 79, a pulsar -- CXOU J185238.6+004020 is discovered to be the central compact object (CCO) of the SNR \citep{Seward2003,Zhou2014}. Two other pulsars -- 3XMM J185246.6+003317 \citep{Zhou2014} and PSR B1849+00 \citep{Cordes2003} are found in the south vicinity of the SNR. About $21.5'$ north from the SNR, there is a more powerful pulsar -- PSR J1853+0056, with a relatively high spin-down power of  $\sim4.1 \times 10^{34}$~erg~s$^{-1}$ \citep{Pellizzoni2002}. 

Pulsar wind nebulae (PWNe) are a notable type of pulsars' derivative products, which form through pulsars' wind materials interacting with and shocked by ambient medium. In terms of energetics, we consider PSR J1853+0056 as a potential candidate for generating a PWN. 

The previous Fermi Large Area Telescope (LAT)   study on Kes 79  detected a point-like GeV source with a significance of $\sim$ 7$\sigma$ \citep{Auchettl2014}. 
This article reports our advanced Fermi-LAT data analysis results on the GeV emissions in/around this complex Kes 79 region, including our discovery of  two spatial components  at $\ge$5~GeV. We then   examine both the hadronic contribution from the SNR \PKHY{(we adopt 7.1~kpc as its distance from us)} and the leptonic contribution from those pulsars. We also discuss the possibility that a putative PWN powered by PSR J1853+0056 could serve as a GeV-bright source. 

\section{Data reduction and analysis for the field around Kes 79}
\label{sec2}
As an improvement on the previous Fermi analysis of Kes 79 by \citep{Auchettl2014}, we have used the newest P8R3 data and newer IRF. In our work, the  Fermitools version 2.0.0 is used to reduce and analyze the Fermi-LAT data. We use 11.5 years of Pass 8 (P8R3) Source class events collected between August 4, 2008, and April 2, 2020. We focus mainly on data at photon energies above 500~MeV for higher angular resolution.  For reference, the LAT point-spread function  is PSF($E$)=3.5$^{\circ}(E/100~MeV)^{-0.8}$ for on-axis photons  \citep{Atwood_Mission_2009, Panaitescu2017}.  The region of interest (ROI) we choose  is $20^{\circ}\times20^{\circ}$ centered at RA =$18^{h}52^{m}48.00^{s}$, Dec=$00^{\circ}40^{'}48.00^{''}$. 

We perform a series of binned maximum-likelihood analyses (with an angular bin size of 0.05$^\circ$ that is sufficiently small to well sample the point-spread function (PSF) at energies up to $\sim$20~GeV). To \PKHY{better model} the background, the diffuse model  {\it gll\_iem\_v07.fits~\footnote{\url{https://fermi.gsfc.nasa.gov/ssc/data/analysis/software/aux/4fgl/gll_iem_v07.fits}}}(Galactic diffuse emission) and {\it iso\_P8R3\_SOURCE\_V2\_v1.txt~\footnote{\url{https://fermi.gsfc.nasa.gov/ssc/data/analysis/software/aux/iso_P8R3_SOURCE_V3_v1.txt}}}(isotropic diffuse component) are used in our analysis, and sources in the Fermi catalog \citep[4FGL;][]{Fermi_Fourth_2019} are included as background sources.  We set free the spectral parameters of the sources within 5$^\circ$ from the ROI center (including the normalizations of the Galactic diffuse background and of the isotropic diffuse component) in each analysis. For the sources at angular separation beyond 5$^\circ$ from the ROI center, their spectral parameters are fixed to the catalog values. 

In the newest 4FGL catalog, 4FGL J1852.4+0037e is associated with SNR Kes 79, and is assigned with the position (18:52:48.00,00:40:48.0) and a \PKHY{disk} radius of 0.1$^{\circ}$  same as those of the SNR  radio shell. In this work, our null hypothesis (to be rejected in \S\ref{sec2.1}) is that the spatial extension of 4FGL J1852.4+0037e (i.e. the SNR radio shell) covers all $\gamma$-ray emissions from the entire SNR Kes 79 region, while our alternative hypothesis (to be affirmed) is that the $\gamma$-rays from this region extend beyond the SNR radio shell and are even decomposable into two components. 

\subsection{Results of spatial analyses }
\label{sec2.1}
\begin{figure*}
\includegraphics[width=90mm,height=80mm]{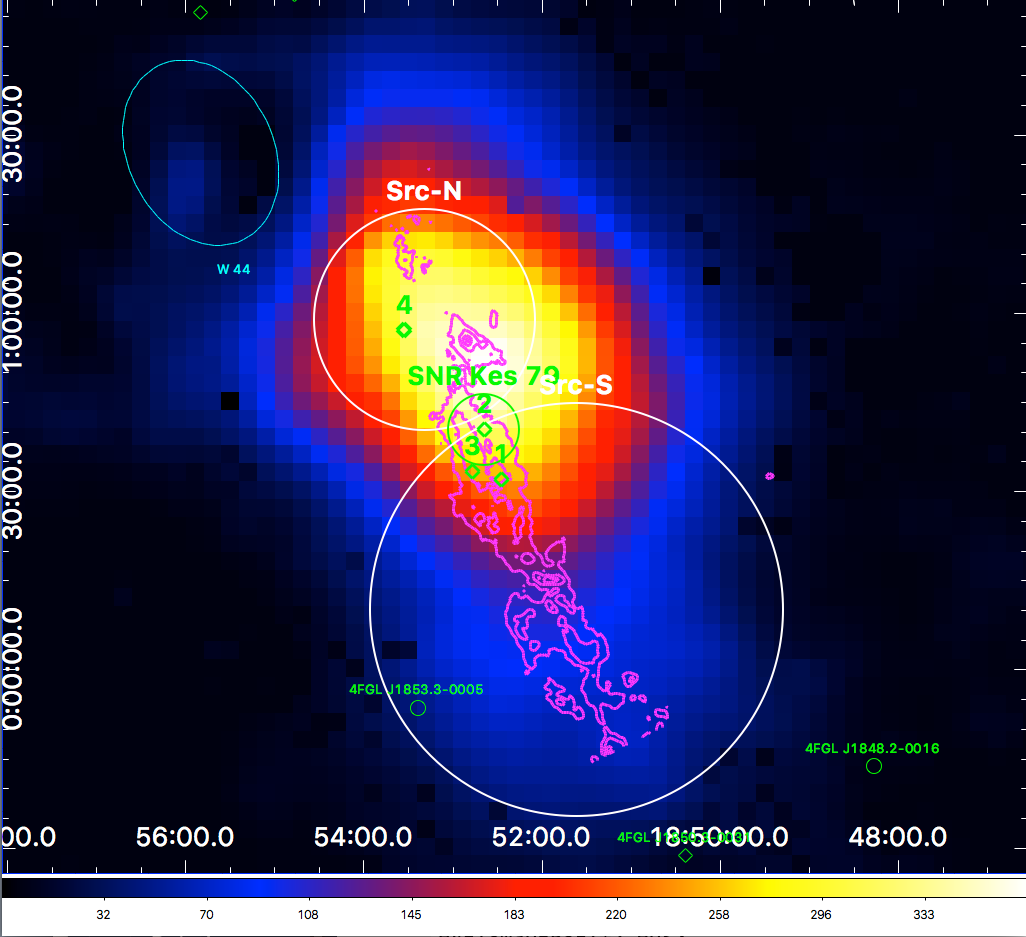}
$\quad$\includegraphics[width=85mm,height=80mm]{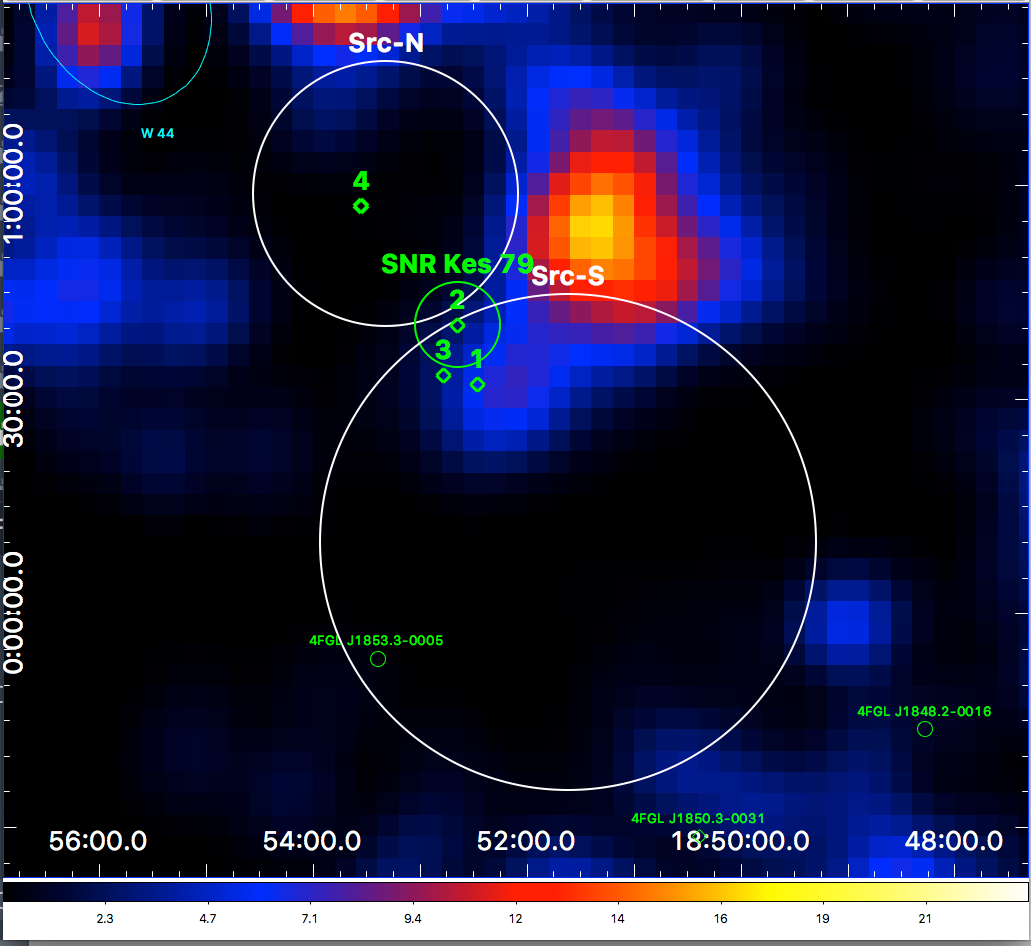}
\caption{Left panel: The 1--50~GeV TS map (for ``FRONT+BACK'' data) of the field around SNR Kes 79, where all neighboring 4FGL catalog sources and diffuse backgrounds are subtracted. It is overlaid with magenta contours of $^{13}$CO $J=1-0$ emission at 110.201~GHz in the velocity range 100--110~km~s$^{-1}$ \citep[BU-FCRAO Galactic Ring Survey data~$^{\ref{GRS}}$;][]{Jackson2006} -- a tracer of MC distribution.  Right panel: The 1--50~GeV residual TS map  where Src-N and Src-S (replacing 4FGL J1852.4+0037e) are modeled out. The white circles marked with ``Src-N" and ``Src-S" represent the best-fit positions of the two sources discovered in our analysis, and the radii of white circles are the most likely extensions of the two sources. The nearby 4FGL sources are marked in green. The green diamonds indicate the position of (1) PSR B1849+00 \citep{Hobbs2004}, (2) CXOU J185238.6+004020 \citep{Seward2003}, (3)3XMM J185246.6+003317 \citep{Zhou2014} and (4) PSR J1853+0056 \citep{Pellizzoni2002}. The position and size of SNR Kes 79 are equal to those of 4FGL J1852.4+0037e. }
\label{sample2}
\end{figure*}

\begin{figure*}
\includegraphics[width=90mm,height=80mm]{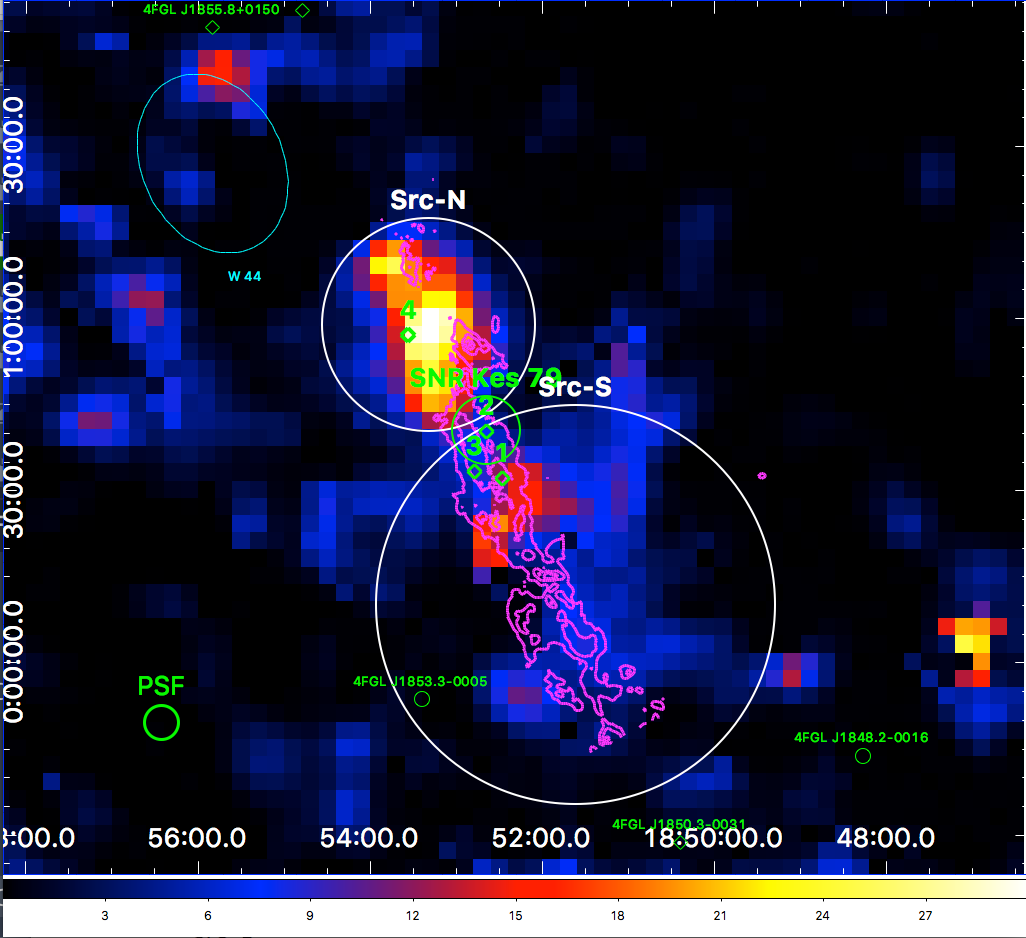}
$\quad$\includegraphics[width=90mm,height=80mm]{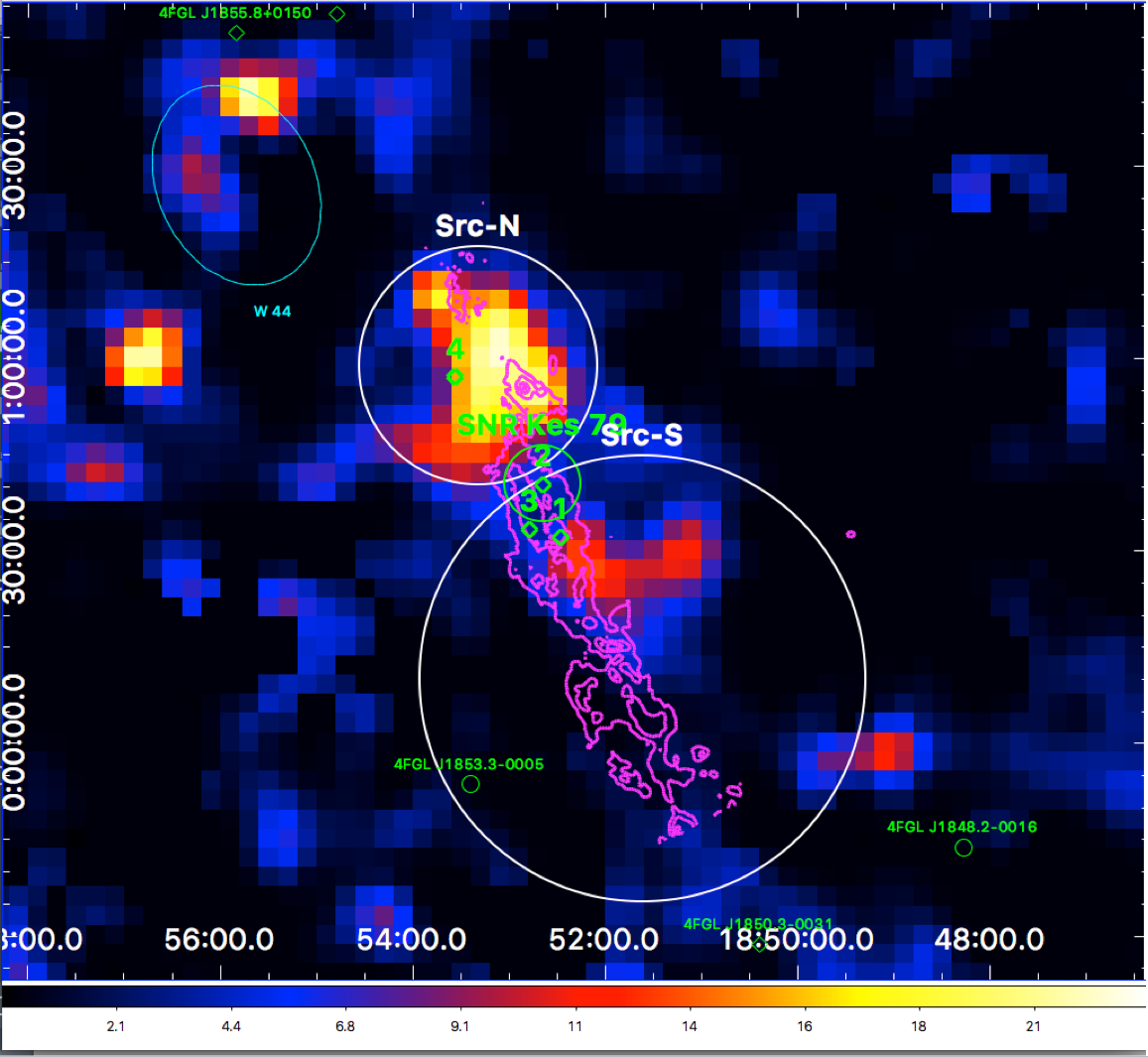}
\caption{The 5--50 GeV TS maps of the  Kes 79 region  for different event types. The left panel is created with ``FRONT+BACK'' data, and the right one is created with ``PSF2+PSF3'' data. The PSF in the left panel is the average point spread function. Both panels are overlaid with magenta contours of $^{13}$CO $J=1-0$ emission at 110.201~GHz in the velocity range 100--110~km~s$^{-1}$ \citep[BU-FCRAO Galactic Ring Survey data~$^{\ref{GRS}}$;][]{Jackson2006} -- a tracer of MC distribution.   The white circles marked with ``Src-N" and ``Src-S" represent the best-fit positions of the two sources discovered in our analysis, and the radii of white circles are the most likely extensions of the two sources. The nearby 4FGL sources are marked in green. The green diamonds indicate the position of (1) PSR B1849+00 \citep{Hobbs2004}, (2) CXOU J185238.6+004020 \citep{Seward2003}, (3)3XMM J185246.6+003317 \citep{Zhou2014} and (4) PSR J1853+0056 \citep{Pellizzoni2002}. The position and size of SNR Kes 79 are equal to those of 4FGL J1852.4+0037e. }
\label{sample3}
\end{figure*}

For examining the morphology of the emission from this region, we adopt the TS maps with different energy ranges. The TS map of the  Kes 79 region in 1--50~GeV for ``FRONT+BACK" data is shown in Figure~\ref{sample2}. On this map, the centroid position is northward offset from SNR Kes 79 (4FGL J1852.4+0037e) by $\sim0.2^\circ$, and the feature is elongated along the molecular-cloud (MC) structure traced by CO emission  \citep[BU-FCRAO Galactic Ring Survey data~\footnote{\label{GRS}\url{https://www.bu.edu/galacticring/new_data.html}};][]{Jackson2006}. \PKHY{We choose the velocity range 100--110~km~s$^{-1}$ for the CO emission contours in Figure~\ref{sample2}, based on the fact that the MCs associated with Kes 79 are located there.} The spatial analysis at higher energies shows that the GeV emission is resolved into two structures. According to the 5--50~GeV TS maps (Figure~\ref{sample3}), we  found a bright emission region (Src-N) at the north of 4FGL J1852.4+0037e. There is no strong  emission around the 4FGL J1852.4+0037e. At the south of 4FGL J1852.4+0037e, there is another clump of emission (Src-S) which is relatively fainter than Src-N. 

\begin{figure*}
\includegraphics[width=135mm,height=90mm]{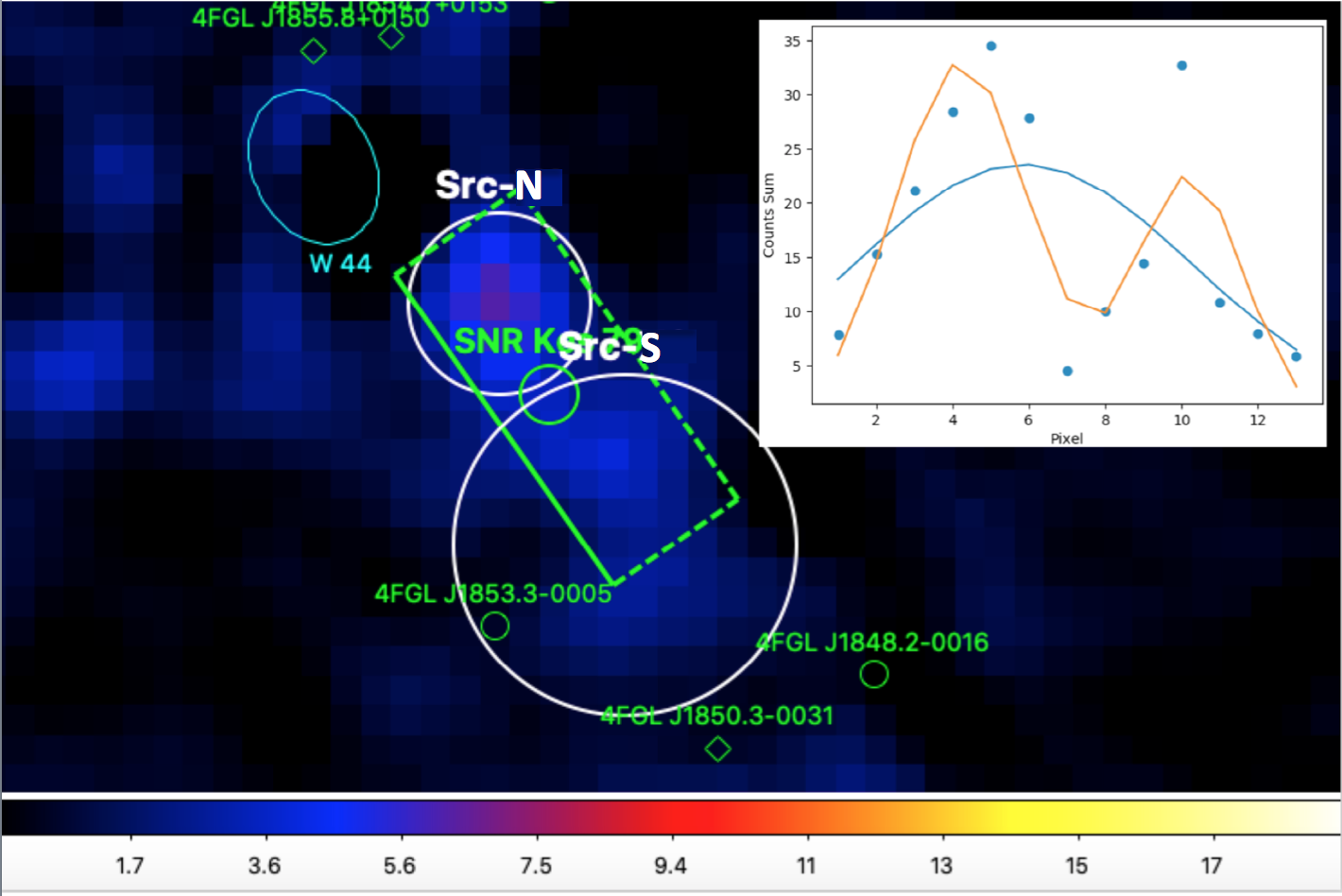}
\caption{A 5--50~GeV count-map (for ``FRONT+BACK'' data) where all neighbouring 4FGL catalog sources and diffuse backgrounds are subtracted, and a brightness profile  computed along Src-N and Src-S. The box in the inset illustrates the orientation that the brightness profile is computed. The orange curve shows the preferable double-Gaussian additive model. The blue curve shows the disfavoured single-Gaussian model. Both models are fitted by maximising the Poissonian log-likelihood function.}
\label{sample-profile}
\end{figure*}

We produced a 5--50~GeV count-map (for ``FRONT+BACK'' data) where all 4FGL catalog sources and diffuse backgrounds are subtracted, and then computed a brightness profile along Src-N and Src-S (Figure~\ref{sample-profile}). In order to confirm the existence of two spatial components, we performed Poissonian log-likelihood fits to the count distribution, with single-Gaussian and double-Gaussian (additive) models respectively. It turns out that a double Gaussian is preferred over a single Gaussian with $\Delta TS = 33.7$ for 3 d.o.f. $(\sim 5.2\sigma)$. Therefore, we claim a significant detection of two sources Src-N and Src-S in the field around Kes 79. 

Then, we use maximum-likelihood analysis to investigate the  respective morphologies of Src-N and Src-S.  In this procedure, we perform joint analyses with ``PSF2" and ``PSF3" data, so as to achieve a compromise between good spatial resolution and adequate photon statistics.  We determine the spatial properties of Src-N (the distinctively brightest component in this region) with 5--50~GeV data for higher spatial resolution. However, we determine those of Src-S with 1--5~GeV data for higher photon statistics, because it has a relatively poorer significance of detection in 5--50~GeV. When looking into Src-S, we model out Src-N with \PKHY{fixing its spatial parameters at our determined position and extension but leaving its spectral parameters free to vary}. We performed a likelihood ratio test to quantify the significance of extension. We assigned power-law spectral models to  Src-N and Src-S, and we attempted uniform-disk morphologies of different radii as well as a point-source model on them.

The best-fit position of Src-N is found to be R.A.=283.33$\degr$, Dec.=0.98$\degr$ in 5--50~GeV, and the best-fit position is R.A.=282.90$\degr$, Dec.=0.17$\degr$ for Src-S in 1--5~GeV. The distance between Src-N and Src-S is larger than the PSF ($<0.55^{\circ}$) in 1--50~GeV. For Src-N,  the 2$\Delta$ln(likelihood) of different radii relative to the point-source model are tabulated in Table~\ref{table1}. The most likely radius is determined to be $0.31\degr\pm0.07\degr$, and this morphology is preferred over a point-source model by $\sim4.3\sigma$. This radius is at least twice the PSF for the $\ge$5~GeV data, justifying the measurement. For Src-S, the most likely radius is determined to be $0.58\degr\pm0.06\degr$, and this morphology is preferred over a point-source model by $\sim6.7\sigma$ (Table~\ref{table2}). This radius is  larger than the PSF for the $\ge$1~GeV data, justifying the measurement. We conclude that the $\ge$ 5~GeV emission is mostly concentrated in two extended sources -- one located to the north of the SNR (Src-N, 6.4$\sigma$ detection) and the other to the south of the SNR (Src-S, 3.9$\sigma$ detection). The positions and extension sizes of Src-N and Src-S are overlaid in Figures~\ref{sample2},~\ref{sample3}~\&~\ref{sample-profile}.  

\begin{table}
\caption{The 2$\Delta$ln(likelihood) in 5--50 GeV for “PSF2+PSF3” data, when uniform disks of different radii replace the point-source model to be the morphology of Src-N.}
\label{table1}
\begin{tabular}{lc}
\hline \hline
 Radius of extension ($\degr$)  \qquad \qquad & \qquad 2$\Delta$ln(likelihood)\\ 
 \hline
0.09  \qquad \qquad & \qquad 4.94\\
0.15  \qquad \qquad & \qquad 10.24\\
0.21  \qquad \qquad & \qquad 16.24\\
0.27  \qquad \qquad & \qquad 17.96\\
0.29  \qquad \qquad & \qquad 18.06\\
0.30  \qquad \qquad & \qquad 18.14\\
0.31  \qquad \qquad & \qquad 18.22\\
0.33  \qquad \qquad & \qquad 18.18\\
0.39  \qquad \qquad & \qquad 17.05\\
0.45  \qquad \qquad & \qquad 13.28\\
\hline
\end{tabular}
\end{table}

\begin{table}
\caption{The 2$\Delta$ln(likelihood) in 1--5 GeV for “PSF2+PSF3” data, when uniform disks of different radii replace the point-source model to be the morphology of Src-S.}
\label{table2}
\begin{tabular}{lc}
\hline \hline
 Radius of extension ($\degr$)  \qquad \qquad & \qquad 2$\Delta$ln(likelihood)\\ 
 \hline
0.10  \qquad \qquad & \qquad 2.41\\
0.20  \qquad \qquad & \qquad 9.68\\
0.40  \qquad \qquad & \qquad 33.81\\
0.55  \qquad \qquad & \qquad 44.11\\
0.57  \qquad \qquad & \qquad 44.39\\
0.58  \qquad \qquad & \qquad 44.44\\
0.59  \qquad \qquad & \qquad 44.41\\
0.61  \qquad \qquad & \qquad 44.14\\
0.65  \qquad \qquad & \qquad 42.75\\
0.70  \qquad \qquad & \qquad 29.38\\
\hline
\end{tabular}
\end{table}

Although it is hard to distinguish any sub-features around the SNR at 1~GeV (PSF$\sim0.55^{\circ}$), we cannot exclude the possibility that the observed whole 1--50~GeV feature may also be a superposition of Src-N and Src-S convolved with the large PSF. In the 1--50~GeV residual TS map (Figure~\ref{sample2}) where Src-N and Src-S replace 4FGL J1852.4+0037e (i.e. the SNR radio shell) to be modeled out, almost nothing is left in the whole field  (the maximum TS value is only $\sim$20, which is tiny compared to the peak of $\sim$380 on the TS map showing the entire feature). We, thus, focus mainly on spectral fittings for Src-N and Src-S in \S\ref{sec2.2}.

\PKHY{The last procedure of our spatial analyses is to quantify the systematic uncertainties of their extension radii stemming from the background subtraction. We cross-check the extensions of Src-N and Src-S by shifting the Galactic diffuse model's normalisation by $\pm$5\%. After reverting the Galactic diffuse model's normalisation to the best-fit values, we also cross-check the Src-N extension by changing the W44 model's normalisation by $\pm$10\%, and cross-check the Src-S extension by changing the Src-N radius by $\pm0.09\degr$. We thereby determine the disk radius of Src-N to be $0.31\degr\pm0.07\degr_{stat}\pm0.06\degr_{sys}$, and that of Src-S to be $0.58\degr\pm0.06\degr_{stat}\pm0.05\degr_{sys}$. }

\subsection{Results of spectral  analyses }
\label{sec2.2}

``FRONT+BACK" data are used to investigate the spectra.
For spectral fittings,  we attempt a  power-law (PL) model: 
\begin{equation}
\frac{dN}{dE} = N_0 \left(\frac{E}{E_0}\right)^{-\Gamma},
\end{equation}
a power-law with an exponential cutoff (PLE):
\begin{equation}
\frac{dN}{dE} = N_0 \left(\frac{E}{E_0}\right)^{-\gamma}exp\left(-\frac{E}{E_{cut}}\right),
\end{equation}
and a broken-power-law (BPL) model:
\begin{equation}
\frac{dN}{dE} = N_\mathrm{0} \left\{
\begin{array}{ll}
	\left(\frac{E}{E_{break}}\right)^{-\alpha}\,\mathrm{if}\,E<E_\mathrm{b}  \\
	\left(\frac{E}{E_{break}}\right)^{-\beta}\,\mathrm{if}\,E\geq E_\mathrm{b} 
\end{array}
\right. .
\end{equation}

According to the likelihood ratio test on  4FGL J1852.4+0037e (i.e. the SNR radio shell), its 0.1--50~GeV spectrum  is best described by BPL which is preferred over PL and PLE with $\Delta TS > 27$.  BPL yields a photon index $\alpha=1.24\pm 0.16$ below the spectral break $E_{break}=497.79\pm12.45$~MeV and $\beta=2.55\pm0.04$ above the break. 4FGL J1852.4+0037e assigned with this BPL is detected with a TS value of 1213.36 ($\sim$34.8$\sigma$ significance) and an integrated flux of (92.0$\pm$9.8)$\times10^{-9}$~photons~cm$^{-2}$~s$^{-1}$ in 0.1--50~GeV.

\begin{figure*}
\includegraphics[width=0.5\linewidth]{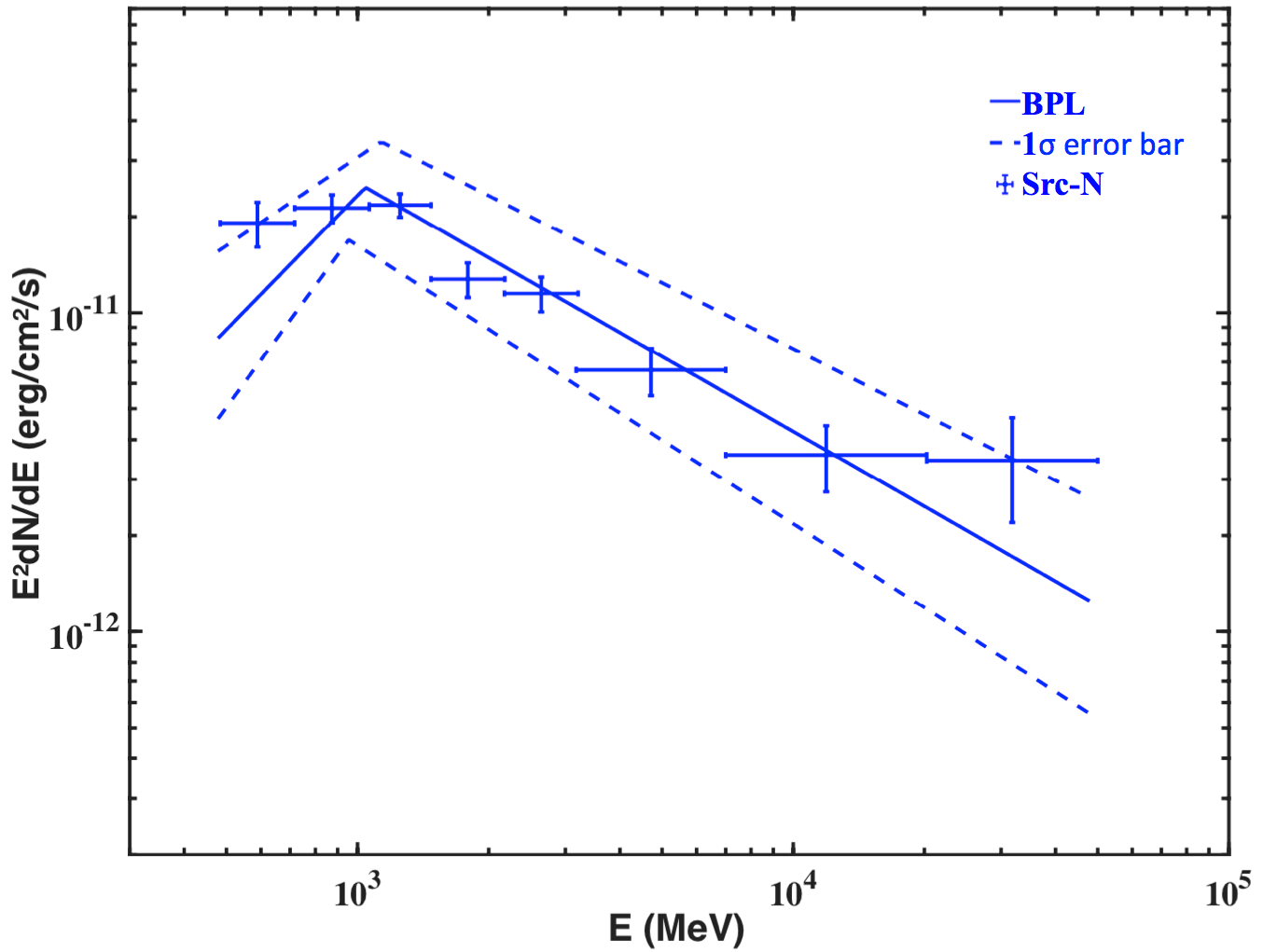}
\includegraphics[width=0.5\linewidth]{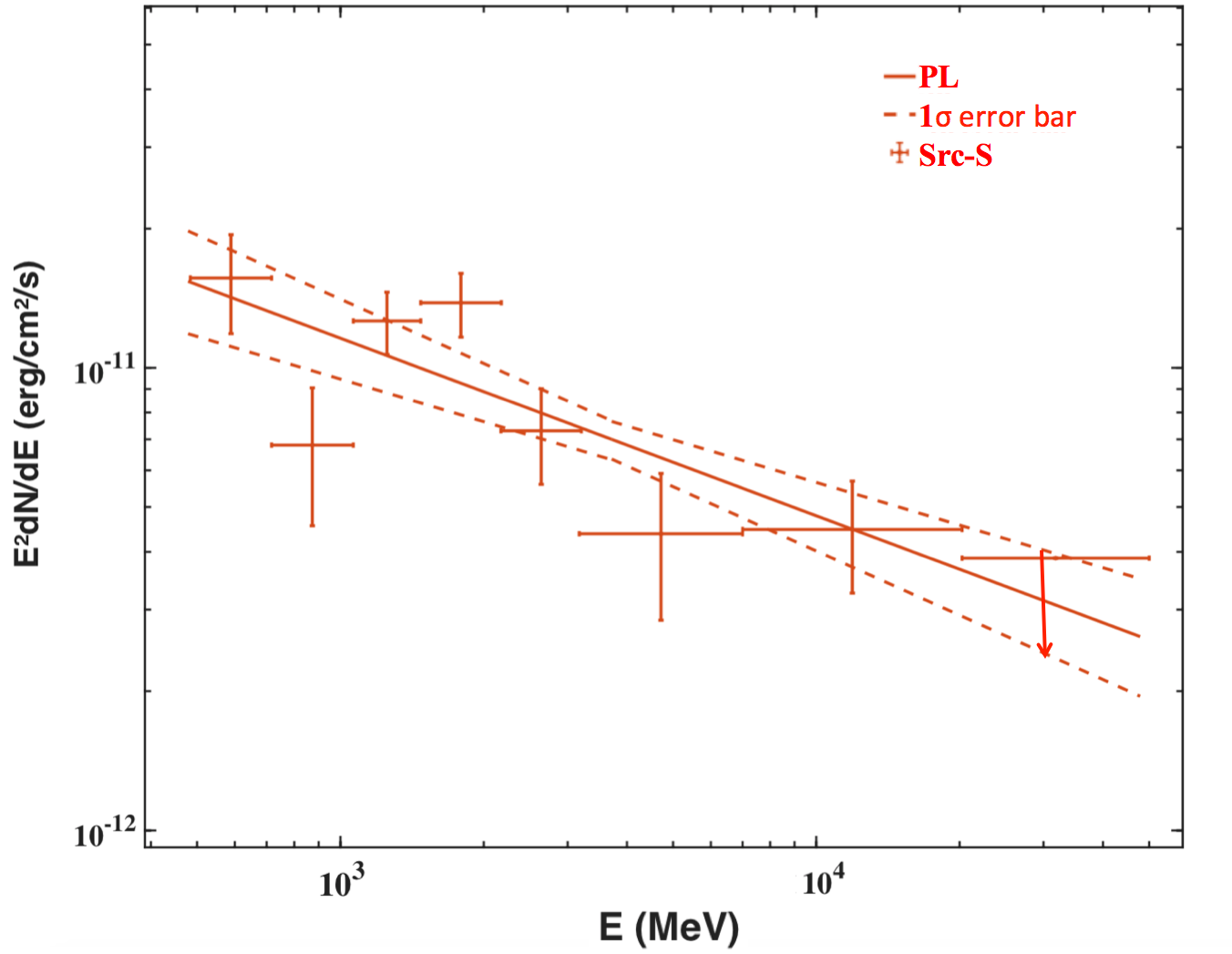}\\
\includegraphics[width=\linewidth]{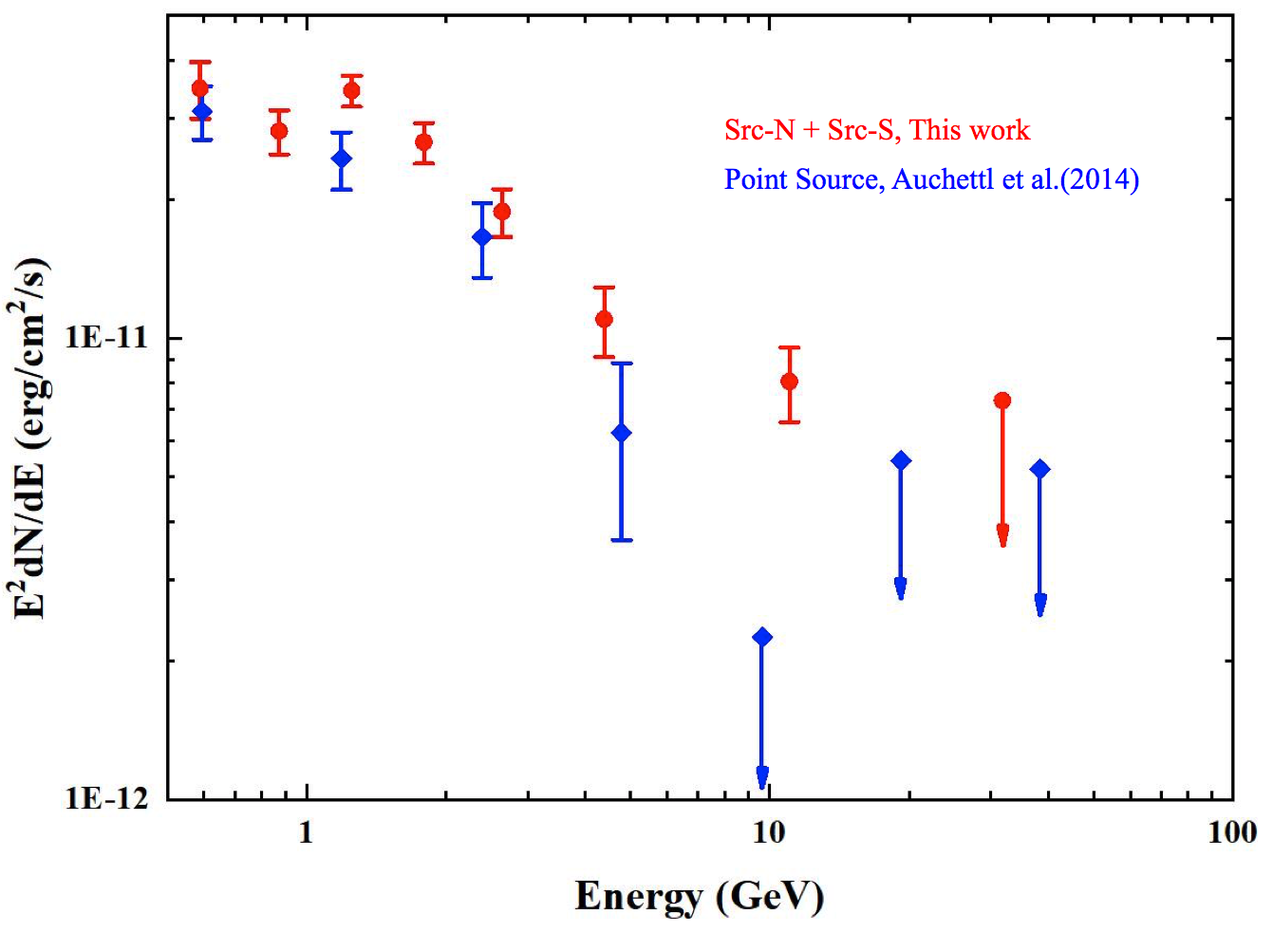}
\caption{The 0.5--50~GeV spectra of Src-N (top-left), Src-S (top-right) and their sum (bottom, in comparison with the point source reported by \citet{Auchettl2014}).  Upper limits are calculated for bins with  TS $<$ 9, fixing the photon index at $\Gamma=3$. }
\label{sample4}
\end{figure*}

Subsequently, in the source model, we replace 4FGL J1852.4+0037e with our defined Src-N and Src-S. In the spectral fittings for them, we exclude the 0.1--0.5~GeV data for avoiding the severe source confusion of photons from Src-N and from Src-S. In view of a strong correlation between the spectral measurements on them, we attempt $3\times3$ combinations of spectral models for them  so as to achieve unbiased analyses.   The 0.5--50~GeV spectral properties of Src-N and Src-S are listed in Tables~\ref{table4}~\&~\ref{table5}, and their spectral energy distributions are shown in Figure~\ref{sample4}. 

\begin{table}
\caption{The sum of TS values of Src-N and Src-S for 500~MeV--50~GeV, with various combinations of spectral models.}
\label{table4}
\begin{tabular}{c|c|ccc}
\hline \hline
&&\multicolumn{3}{c}{Src-N} \\
\hline
&  & PL & PLE  & BPL\\ 
\hline
&PL & 829.4 & 853.0  & 898.6 \\
Src-S&PLE & 846.9  & 858.7 & 899.3 \\
&BPL & 860.3  & 854.8 & 905.8 \\
\hline

\end{tabular}
\end{table}

\begin{table}
\caption{The spectral parameters  of Src-N and Src-S in 0.5--50~GeV.}
\label{table5}
\begin{tabular}{lcc}
\hline 
  &Src-N & Src-S\\
\hline
\multicolumn{3}{c}{PL model} \\
 \hline
$\Gamma$ &  2.43 $\pm$ 0.04  ($\pm$ 0.05)  & 2.39 $\pm$ 0.07 ($\pm$ 0.13) \\
Flux ($10^{-9}$~photons~cm$^{-2}$~s$^{-1}$) &19.2 $\pm$ 0.9  ($\pm$ 4.7) &  13.5 $\pm$ 0.9  ($\pm$ 7.0) \\
\hline 
\multicolumn{3}{c}{BPL model} \\
 \hline
$\alpha$  & 0.59 $\pm$ 0.45 ($^{+0.48}_{-1.20}$) & 3.36 $\pm$ 0.80 ($\pm$ 0.24) \\
$\beta$  &2.78 $\pm$ 0.09  ($\pm$ 0.02) &  2.30 $\pm$ 0.10 ($\pm$ 0.12) \\
$E_{break}$ (MeV)  &1041.0 $\pm$ 86.9  ($\pm$ 19.0) &  793.6 $\pm$ 249.6 ($\pm$ 5.5) \\
Flux ($10^{-9}$~photons~cm$^{-2}$~s$^{-1}$)  &17.6 $\pm$ 1.0 ($\pm$ 5.5) &  13.8 $\pm$ 1.0  ($\pm$ 7.0) \\
 \hline
\end{tabular}

	\raggedright
\bf{Note.} \PKHY{The first uncertainties are statistical, and the second uncertainties (i.e., those inside brackets) are systematic. } \\
\end{table}

Src-N is best described by a BPL model, which is preferred over PL and PLE with $\Delta TS > 40$. BPL of Src-N yields a photon index $\alpha=0.59\pm0.45$ below the spectral break (i.e. the peak of the differential energy flux) $E_{break}=1041.0\pm86.9$~MeV and $\beta=2.78\pm0.09$ above the peak. For Src-S, a PL model with a photon index $\Gamma=2.39\pm0.07$ is sufficient to fit its 0.5--50~GeV spectrum, and the extra parameters of PLE or BPL are not strongly required (only $\le2.2\sigma$). More importantly, even in the BPL model of Src-S, the photon index $\alpha$ below the spectral break appears to be steeper than the photon index $\beta$ above the break, entailing that the peak of the differential energy flux of Src-S lies at an energy $\lesssim$0.5~GeV.

\PKHY{We proceed to quantify the systematic uncertainties of their spectral parameters stemming from the background subtraction. To cross-check the spectral properties of Src-N and Src-S, we alter the Galactic diffuse model's normalisation by $\pm$5\% and alter the W44 model's normalisation by $\pm$10\%. Each of them is changed while the other is fixed at its best-fit value. The determined systematic uncertainties are appended in Table~\ref{table5} (inside brackets). For all parameters except the integrated flux, the average systematic uncertainty is smaller than a double of the corresponding statistical uncertainty. Noteworthily, our detection of the Src-N’s spectral peak at $\sim$1.0~GeV is robust against both the systematic effect of background subtraction and the statistical fluctuations. }

\section{Discussion}

From our analysis, the $\gamma$-rays from the SNR Kes 79 region are robustly detected by Fermi LAT ($\sim$34.8$\sigma$ in 0.1--50~GeV). While \citet{Auchettl2014} detected and modelled the whole $\gamma$-ray feature as a single point source, we found that the $\ge$ 5~GeV emission is mostly concentrated in two extended sources -- a brighter one located to the north of the SNR (Src-N) and a relatively fainter one to the south of the SNR (Src-S). 

As demonstrated in the bottom panel of Figure~\ref{sample4}, the total flux of Src-N and Src-S appears to slightly outweigh the flux of the point source reported by \citet{Auchettl2014} in the 0.5--7~GeV energy range, and is significantly higher than that at energies above 7~GeV. These excessive fluxes, as well as the hardened spectrum, are explainable by the large coverage  of the extended-source models of Src-N and Src-S. 

The spectral fitting  by \citet{Auchettl2014} yields a preferable PLE model with a cutoff energy of 2.71 $\pm$ 0.64 GeV. Differently, our spectral analysis in 0.5--50~GeV indicates that a BPL model best describes the Src-N spectrum, while \PKHY{PLE and BPL  cannot significantly improve the likelihood over PL for} the Src-S spectrum. Their distinct spectral shapes suggest that they may have different origins. 

\subsection{Is Src-N powered by SNR-MC interactions?}
\label{sec3.1}

As shown in Figure~\ref{sample2}, SNR Kes 79 is surrounded by  giant MCs traced by CO emission, which include those inside Src-N. These giant MCs  provide  protons for hadronic interactions. SNRs can inject the particles accelerated through diffusive shock acceleration (DSA) into the nearby clouds. There are two scenarios for releasing CRs: (i) The CRs can run away from the shock surface when the particles are accelerated above the escape energy ($E_{max}$); (ii) The GeV CRs can be leaked from a broken shell after an SNR-cloud collision \citep{Cui2018}. 

According to \citet{Zhou2016}, SNR Kes 79 is a mid-aged SNR with Sedov age $\sim$ 4.4-6.7 kyr and \PKHY{the current shock  velocity is $u_{sh}\sim$730~km~s$^{-1}$ with the shock radius $R_{sh}\sim 6^{\prime}$ (corresponding to $\sim$12~pc at a 7.1~kpc distance). They estimate the supernova explosion kinetic energy to be ${\mathcal{E}_{51}}\sim2.7\times10^{50}\,$erg and the pre-shock gas density to be $n_0 \sim 0.13 cm^{-3}$. This value of $n_0$ implies that the shock is expanding in a stellar wind bubble launched by a massive progenitor. Thermal X-ray emission together with infrared observations by \citet{Zhou2016} has justified that the shock has been colliding with surrounding materials and may have resulted in a broken shell (i.e., the aforementioned scenario (ii) is viable for Kes 79). Due to the low shock velocity, it is not surprising that \citet{Zhou2016} did not detect any non-thermal X-ray emission at the SNR, and the non-detection of such X-ray synchrotron emission normally marks its failure to be a TeV electron accelerator. On the other hand, we cannot exclude the possibility that Kes 79 is still capable of accelerating protons up to super-TeV. Adopting the non-resonant acceleration theory \citep{Zirakashvili2008} and the equation (7) of \citet{Ptuskin2005}, we derive a current escape energy of $E_{max}\geq$1~TeV (for the scenario (i)) with an Alfv\'en velocity $V_A\sim$10~km~s$^{-1}$ \citep{Zirakashvili2008} and an magnetic field $B_0$=5~$\mu$G \citep{Ptuskin2005} in the circumstellar medium.} 

Nevertheless, Kes 79 is not detected (only $\sim 2.7\sigma$ with observation time $\sim$51.4h) in the TeV band by HESS \citep[cf.][]{Bochow2011ASS}. Therefore, this work focuses on SNR-cloud interactions (the scenario (ii)) which can easily release CRs down to several GeV. Most of the accelerated CRs with energy below $E_{max}$ are confined in the downstream not far from the shock front. After the collision with the dense clouds, the shock is quickly stalled and can no longer maintain strong magnetic turbulence to confine those GeV CRs  \citep{Zirakashvili2008}. We assume that the energy distribution of the leaked CRs from the broken shell follows a power-law with an exponential cutoff:
\begin{equation}
F_{CR} = N_{CR} \left(\frac{E}{E_0}\right)^{-2}exp(-\frac{E}{1~\mathrm{TeV}})
\end{equation}
At diffusion time $t_{d}$ after the CRs are released from the shock, the final CR density at distance $R_{d}$ from the shock can be simply written as:
\[
n_{\mathrm{R_{d}}}(E,R_{d},t_{d}) = F_{CR}G(E,R_{d},t_{d}),
\]
where $G(E,R_{d},t_{d})=1/8(\pi t_{d} D)^{-3/2} \exp[-R_{d}^2/(4 t_{d}D)]$,  $\int_0^\infty dR_{d} 4\pi R_{d}^2  G=1$ and $D(E)$ is the energy-dependent diffusion coefficient \citep{Thoudam2007}. 

In our proton-proton (pp) interaction model, constraints on some parameters for modelling are explained as follows:

$\bullet  \quad$Canonically, a typical supernova explosion releases kinetic energy of $ {\mathcal{E}_{51}}\sim10^{51}\,$erg.  In our pp model, we adopt this typical $ {\mathcal{E}_{51}}\sim10^{51}\,$erg (about a quadruple of the value estimated for Kes 79 by \citet{Zhou2016}) for systematic analysis.  To calculate the CR energy, we  assume an efficiency of $\eta\sim10\%$ for converting the shock energy to non-thermal CR energy \citep{Ginzburg1964}.

$\bullet  \quad$After supernova explosion, the shock expand in the stellar wind bubble. Sometimes, the shock encounters with those dense clouds which survived the stellar wind. In our model, we give extreme value boundaries for $X=0-50\%$, where X is the ratio between stalled shock area by dense clouds and the total shock area. 

$\bullet  \quad$ The projected distance between the center of SNR and the brightest location of Src-N is $R_{d}\sim$36~pc ($\sim0.3^{\circ}$). According to the 5--50~GeV TS map and the PSF of Fermi-LAT, the $>$5~GeV emission at the radio shell of SNR Kes 79 is much fainter than that at Src-N, so we consider that the diffusion distance should be larger than the scale of SNR Kes 79. We assume an average $R_{d}$ $\sim$36~pc for the diffusion distance. SNR Kes 79 is a mid-aged SNR with a Sedov age $\sim$4.4--6.7~kyr. Thus, we assume the diffusion time is $t_{d}$=1--7~kyr in our model. 

$\bullet  \quad$The diffusion coefficient is usually assumed to be a power-law function of energy, $D(E) = D_{10}(E/(10~GeV))^{\delta}$. The typical values \PKHY{for interstellar medium} are $D_{10}=10^{28}cm^2/s$ and $\delta= 0.5$. In our model, we fit the spectrum with ranges of $D_{10}$ and $\delta$ values, \PKHY{taking into consideration that the diffusion coefficient could be much smaller in the vicinity of SNRs or inside MCs}.

$\bullet  \quad$The Src-N region harbours gaint MCs, whose total mass  is $\sim67000M_{\odot}$.

$\bullet  \quad$We adopt the cross-sections ($\sigma_{pp}$) $\sim$ 40 mb of pp interactions from \citet{Kelner2006}. A collision between two protons has a one-third probability for generating a neutral pion which quickly decays into two gamma-ray photons.

\begin{figure*}
\includegraphics[width=90mm,height=70mm]{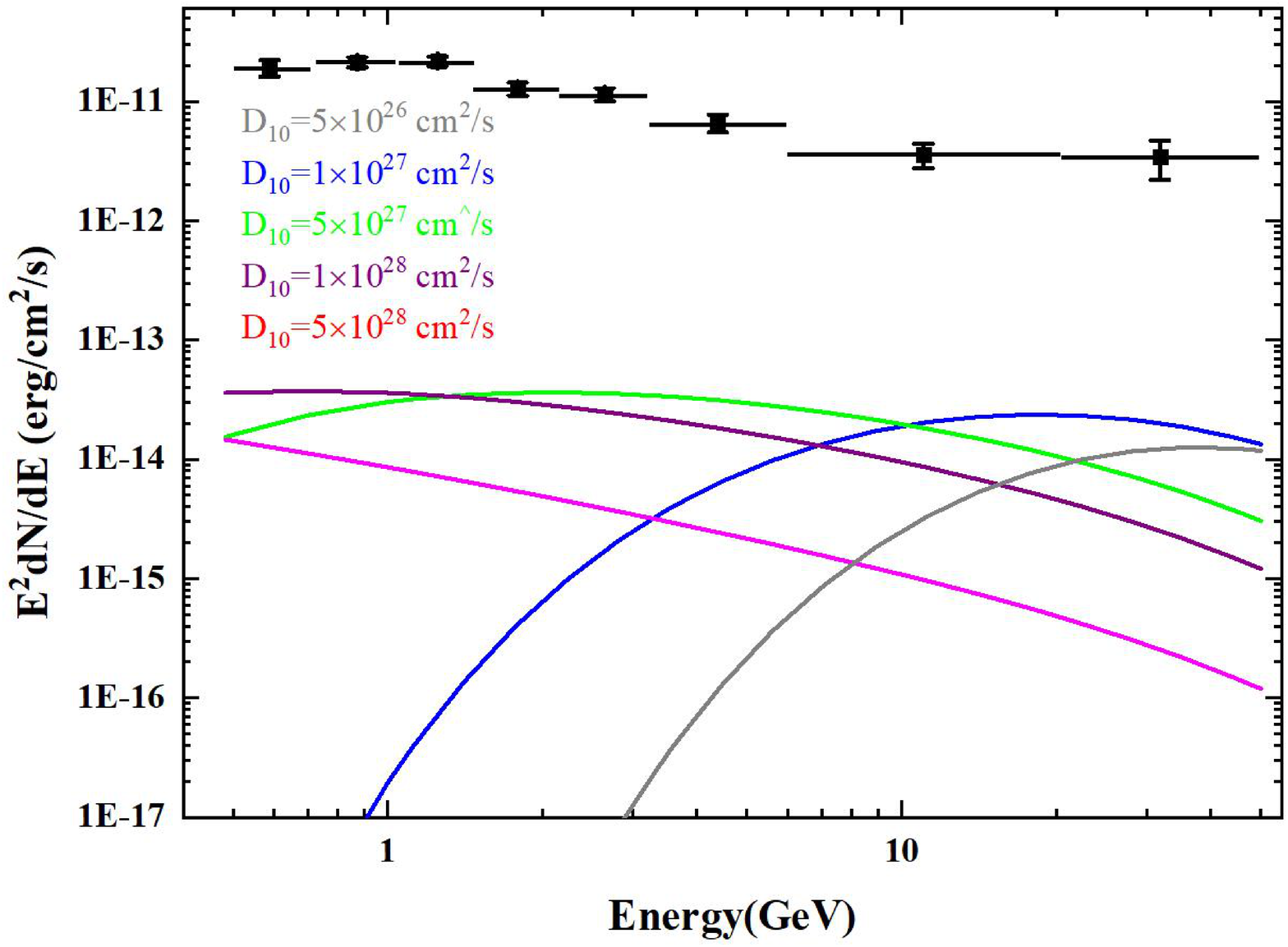}
\includegraphics[width=90mm,height=70mm]{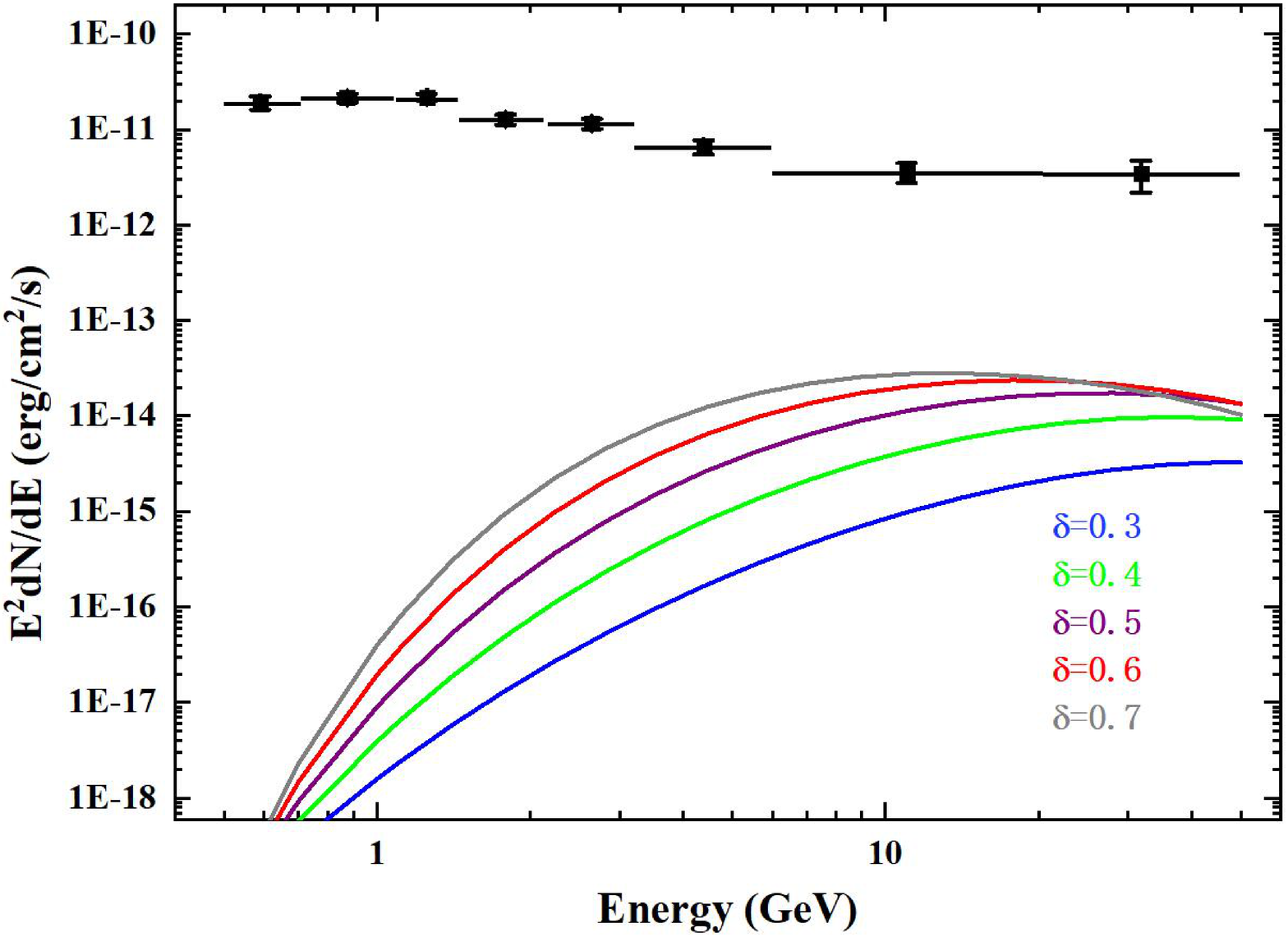}
\includegraphics[width=90mm,height=70mm]{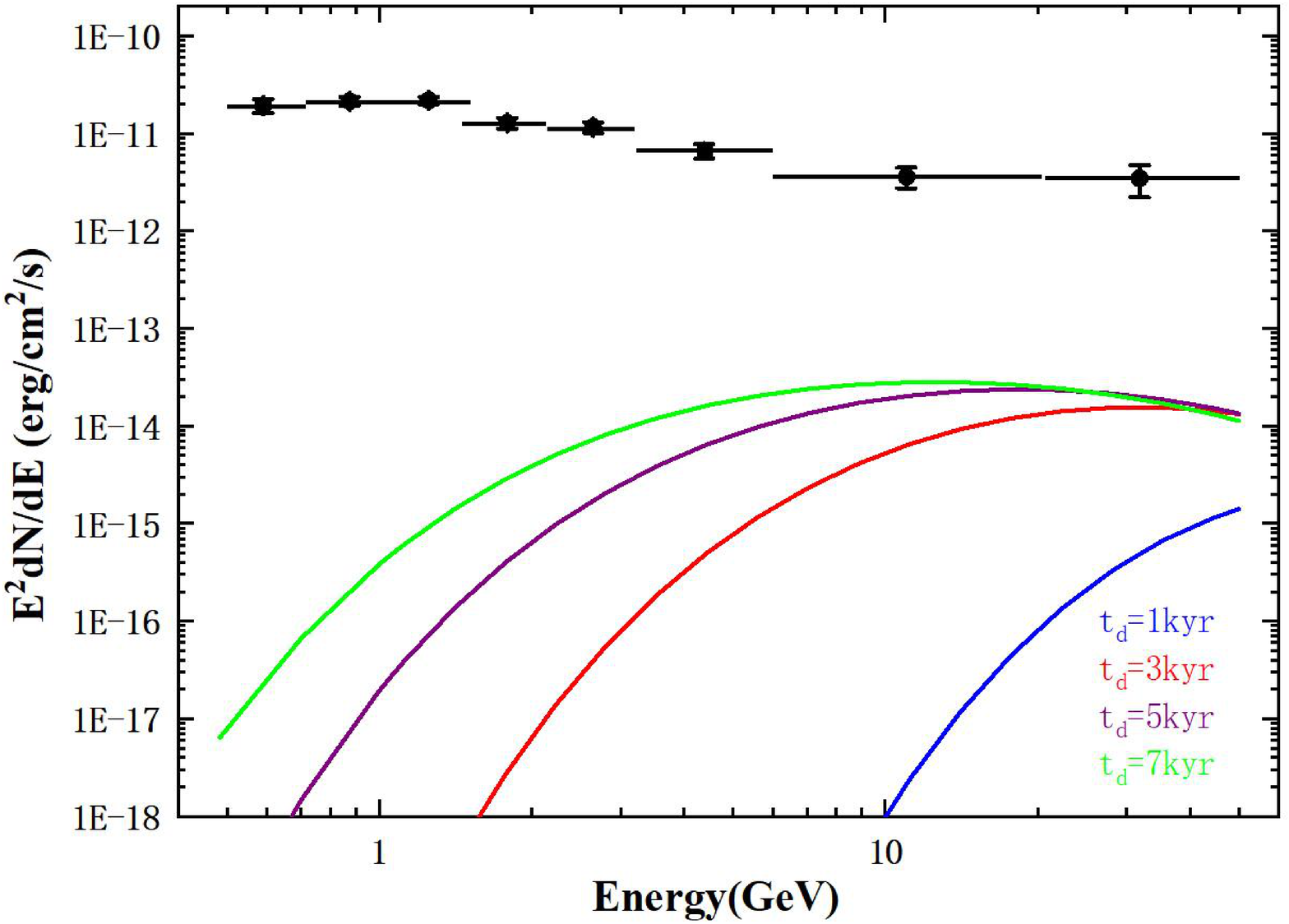}
\caption{The hadronic model fitting for the spectrum of Src-N, assuming that SNR Kes 79 is the only provider of CR protons. Top left panel: the $X=50\%$, $\delta$=0.6 and diffusion time $t_{d} = 5~kyr$. Top right panel: the $X=50\%$, $D_{10}=10^{27}~cm^2/s$ and diffusion time $t_{d} = 5~kyr$. Bottom panel: the $X=50\%$, $D_{10}=10^{27}~cm^2/s$ and $\delta$=0.6.}
\label{sample5}
\end{figure*}

We push the values of the above parameters into extreme, by choosing $X=50\%$, $R_{d}=36~pc$, as illustrated in Figure~\ref{sample5}. However, the hadronic model  predicts insufficient  CRs at Src-N  and fails  to explain the local GeV spectrum. We infer that the relative contribution of SNR Kes 79 to the observed Src-N emission is only $<3\%(\frac{\mathcal{E}_{51}}{10^{51}~\mathrm{erg}}\frac{\eta}{0.1}\frac{\sigma_{pp}}{40~\mathrm{mb}})$ (a conservative upper-limit).

\subsection{Is Src-N related to PSR J1853+0056 or its putative PWN?}
\label{sec3.2}

A powerful pulsar -- PSR J1853+0056 (R.A.=283.39\degr, Dec.=0.95\degr) is located inside the Src-N region.  However, above the spectral break $E_{break}\sim$1.0~GeV, the Src-N spectrum follows a power-law (the photon index $\beta\sim$2.78) which continuously extends to $\gtrsim$30~GeV without an exponential cutoff. Such a spectral shape contrasts with typical $\gamma$-ray spectra of pulsars, which demonstrate exponential cutoffs $E_{cut}$ of about  1--4~GeV \citep{abdo_second_2013}. Worse still, the pulsar itself cannot account for the extended emission (radius $\approx0.31\degr$) of Src-N.

On the other hand, with the reduced synchrotron losses for high-energy inverse-Compton-emitting electrons, a PWN can maintain its high super-GeV  $\gamma$-ray flux  for a timescale exceeding the lifetime of its progenitor pulsar \citep{Tibolla2011}. Such electrons of a PWN are normally believed to be  accelerated  at the terminal shock. 

\begin{figure}
\includegraphics[width=90mm,height=65mm]{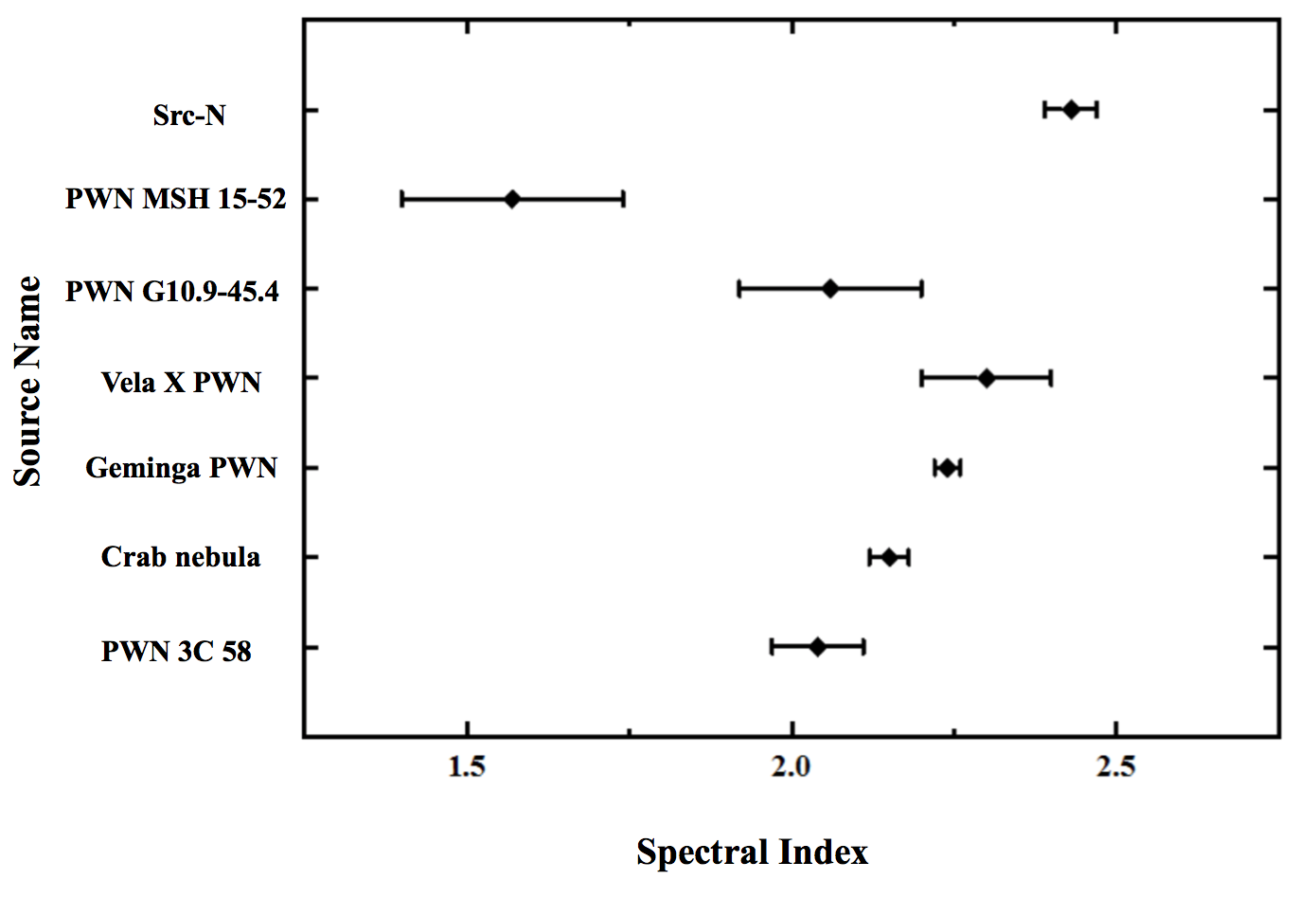}
\caption{The photon spectral index distribution of PWNe. The spectral energy range for  Crab Nebula, Geminga PWN, Vela X PWN, PWN G10.9-45.4 is 0.1--100~GeV \citep{AckermannPWN2011}, that for PWN MSH 15-52 is 1--100~GeV \citep{AbdoPWN2010}, and that for PWN 3C 58 is 0.1--300~GeV \citep{LiPWN2018}.}
\label{PWNcat}
\end{figure}

To further examine the PWN scenario of Src-N, we compared the MeV--GeV spectrum of Src-N with those of other known PWNe. \PKHY{Based on the MeV--GeV spectra of 57 PWNe in the Pulsar Wind Nebula Catalog \citep{roberts2004pwn}, we found that  Crab Nebula, Geminga PWN, Vela X PWN, PWN G10.9-45.4 \citep{AckermannPWN2011}, PWN MSH 15-52 \citep{AbdoPWN2010}, and PWN 3C 58 \citep{LiPWN2018} are significantly detected by Fermi-LAT. We plot the photon indices in Figure~\ref{PWNcat}. According to the photon index distribution,  the Src-N index is relatively high compared to the others. Nevertheless, we still consider it feasible that PSR J1853+0056  powers an unseen PWN which could predominantly reproduce the Src-N emission, based on the energetics elucidated in the following}.

TeV PWNe are generally associated with pulsars releasing power of $>10^{36}$~erg~s$^{-1}$ \citep{Halpern2010b}, but the spin-down power of PSR J1853+0056 \citep[$\sim4.1 \times 10^{34}$~erg~s$^{-1}$;][]{Pellizzoni2002} is lower than this threshold by 1.5 orders of magnitude, making its putative PWN unlikely to be TeV-bright.  Therefore,  the PWN scenario could explain why the Src-N spectrum greatly softens to $\beta\sim$2.78 above $E_{break}\sim$1.0~GeV and is extrapolated to a very low (undetectable) flux in the TeV band. Likewise, the non-detection of a TeV counterpart in this region \citep[$<2.7\sigma$ with observation time $\sim$51.4h  by HESS;][]{Bochow2011ASS} is also explainable in this framework.

\subsection{Relative contribution of SNR-MC interactions  to Src-S}
\label{sec3.3}

Src-S also harbours giant MCs, making it another site of hadronic interactions. On the one hand, spectral peaks are characteristic features of SNRs' hadronic spectra, because the total cross section of inelastic proton-proton collision (mainly via neutral-pion channel) furiously drops when the kinetic energy of the proton is below $\sim$2 GeV \citep[cf. Figure 1 of][]{Kafexhiu2014}. On the other hand, our  fitting with 0.5--50~GeV data does by no means reveal a sharp turnover in the Src-S spectrum. Therefore, we can just place an upper limit of $\lesssim$0.5~GeV on its peak energy.

Noticeably, \citet{Suzuki2018,Suzuki2020a,Suzuki2020b} suggest that the escape of particles from the vicinity of an SNR probably develops with its plasma age, which is positively correlated with the SNR's own age. Their scenario for the SNR evolution specifically predicts that, in the vicinities of mid-aged ($>$3~kyr) SNRs like Kes 79, higher-energy particles generally escape earlier and easier \citep[due to their larger diffusion lengths;][]{Ptuskin2003} than lower-energy particles, causing the lower-energy particles to account for a larger portion of cosmic-ray energy.  Hence, the relatively soft spectrum of Src-S, as well as its low peak energy of $\lesssim$0.5~GeV, is consistent with this scenario if we assume Kes 79 is the major origin of the Src-S emission.

\begin{figure}
\includegraphics[width=90mm,height=70mm]{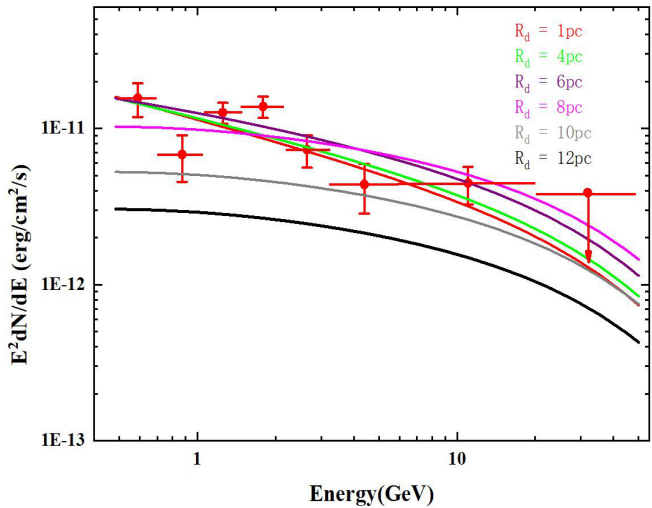}
\caption{The hadronic model fits for the spectrum of Src-S with different diffusion distance $R_{d}$. The $X=50\%$, $\delta$=0.3 and diffusion time $t_{d} = 1~kyr$.  }
\label{X2pp}
\end{figure}

We  adopt the same hadronic  model of Kes 79 as in \S\ref{sec3.1} to describe the spectrum of Src-S. The fitting results are shown in Figure~\ref{X2pp}. The goodness of fit is satisfactory only for a diffusion distance $R_d\le8~pc$. Nevertheless, this diffusion distance corresponds to $\le0.065\degr$ (at 7.1~kpc) which is much smaller than the GeV extension (radius $\approx0.58\degr$) of Src-S. 

Hence, hadronic interactions of Kes 79 with MCs can only dominate a northeast part of Src-S which is closest to the SNR. This northeast part actually contains a majority of the Src-S emission at $\ge$5~GeV, as demonstrated in Figures~\ref{sample3}~\&~\ref{sample-profile}. The excessive extension of Src-S could be attributed to  contamination  by residual Galactic diffuse emission and  systematic uncertainties associated with the PSF (we recall that the Src-S extension is determined with  1--5~GeV data).

\subsection{Relative contribution of pulsars to Src-S}

Inside the Src-S region,  three pulsars are found near the northeast edge. The central compact object (CCO) of Kes 79, CXOU J185238.6+004020 \citep{Seward2003},  manifests itself as an ``anti-magnetar"  with a very low spin-down power and surface magnetic field of $\sim3\times10^{32}$ erg s$^{-1}$ and $\sim3\times10^{10}$   G respectively \citep{Halpern2010a}.  3XMM J185246.6+003317  was discovered to be a transient magnetar \citep[][]{Zhou2014} with a very low spin-down power of $<3.5\times10^{30}$  erg s$^{-1}$ and a relatively lower surface magnetic field of $<4.1\times10^{13}$  G  \citep[both at a $3\sigma$ significance;][]{Rea2014}.  PSR B1849+00 has a spin-down power of  $\sim4\times10^{32}$ erg s$^{-1}$  and a surface magnetic field of  $\sim3\times10^{13}$   G    \citep[cf.][]{Hobbs2004}, located $8.4\pm1.7$  kpc from us \citep[][]{Cordes2003}. 

\PKHY{To examine their relative contribution of $\gamma$-rays, we conservatively assume for each of them a shorter distance $d=7~\mathrm{kpc}$ from us and a narrower beam solid angle $\Omega=3\pi$. It turns out that the total spin-down  flux  of  these three pulsars is only a fraction $\sim5.0\times10^{-3}(\frac{d}{7~\mathrm{kpc}})^{-2}(\frac{\Omega}{3\pi})^{-1}$ of the observed 0.5--50~GeV flux $\sim3.2\times10^{-11}$~erg~cm$^{-2}$~s$^{-1}$ at Src-S.} Even in terms of the power of magnetic field decay \citep[for detail, see][]{Zhang2003}, a  combined  contribution from these three pulsars can hardly supply half of the entire emission of Src-S.

\section{Summary}
We analysed the first 11.5-year Fermi-LAT data for the Kes 79 region with  the newest P8R3 data, the corresponding IRF and most updated source files. Compared with the work of \citet{Auchettl2014}, our result shows a more significant detection ($\sim$34.8$\sigma$ in 0.1--50~GeV) for this region. In addition, we found that the $\ge$5~GeV emission is resolved into two extended sources Src-N (the  brighter one) and Src-S. Their spectra  peak at distinct energies. 

Both a dense MC clump and a relatively powerful pulsar -- PSR J1853+0056 are found inside the region of Src-N. We explore hadronic and leptonic models in attempts to explain the GeV emission  of Src-N. Unfortunately,  the SNR can by no means provide enough hadronic CRs reaching clouds at Src-N to explain the local GeV spectrum. Also, the pulsar itself cannot account for the extension size and spectral shape of Src-N. On the other hand, we propose that a putative PWN powered by PSR J1853+0056 could predominantly reproduce the  observed  $\gamma$-rays at Src-N via leptonic mechanisms, and could account for its dramatic spectral softening above the peak energy $\sim$1.0~GeV as well as its non-detection in TeV by HESS. In the future, detailed X-ray observations of the Src-N region can reveal the concrete emission mechanisms.

The Src-S spectrum has a low peak energy of $\lesssim$0.5~GeV. In terms of the development of particle escape throughout the SNR evolution history, this low peak energy and the relatively soft spectrum are in agreement with  an SNR-MC interaction scenario involving a mid-aged/old ($>$3~kyr) SNR like Kes 79  \citep{Suzuki2018,Suzuki2020a,Suzuki2020b}. By phenomenological fittings, our hadronic model for Kes 79 can  reproduce a majority of the Src-S emission which is closest to the SNR. Three pulsars are also found inside Src-S, but they are too weak to explain half of the observed GeV emission.

\section*{Acknowledgments}
This work is supported by the National Natural Science Foundation of China (NSFC) grants 11633007 and U1731136, Guangdong Major Project of Basic and Applied Basic Research (Grant No. 2019B030302001), Key Laboratory of TianQin Project (Sun Yat-sen University), Ministry of Education, and the China Manned Space Project (No. CMS-CSST-2021-B09). Y.Z. thanks NSFC grant number 11973099 for financial support. Y.C. thanks the NSFC grants 12173018 \& 12121003 for financial support. This work made use of the LAT data and Fermitools available at the Fermi Science Support Center (FSSC). We thank the Help Desk of FSSC for their useful advice on our joint analyses with ``PSF2" and ``PSF3" data.

\begin{thebibliography}{}
\expandafter\ifx\csname natexlab\endcsname\relax\def\natexlab#1{#1}\fi
\providecommand{\url}[1]{\href{#1}{#1}}
\providecommand{\dodoi}[1]{doi:~\href{http://doi.org/#1}{\nolinkurl{#1}}}
\providecommand{\doeprint}[1]{\href{http://ascl.net/#1}{\nolinkurl{http://ascl.net/#1}}}
\providecommand{\doarXiv}[1]{\href{https://arxiv.org/abs/#1}{\nolinkurl{https://arxiv.org/abs/#1}}}

\bibitem[{{Abdo} {et~al.}(2010){Abdo}, {Ackermann}, {Ajello}, {Allafort},
  {Asano}, {Baldini}, {Ballet}, {Barbiellini}, {Baring}, {Bastieri}, {Bechtol},
  {Bellazzini}, {Berenji}, {Blandford}, {Bloom}, {Bonamente}, {Borgland},
  {Bregeon}, {Brez}, {Brigida}, {Bruel}, {Buson}, {Caliandro}, {Cameron},
  {Camilo}, {Caraveo}, {Carrigan}, {Casandjian}, {Cecchi}, {{\c{C}}elik},
  {Chekhtman}, {Cheung}, {Chiang}, {Ciprini}, {Claus}, {Cohen-Tanugi},
  {Conrad}, {den Hartog}, {Dermer}, {de Luca}, {de Palma}, {Dormody}, {Silva},
  {Drell}, {Dubois}, {Dumora}, {Farnier}, {Favuzzi}, {Fegan}, {Ferrara},
  {Focke}, {Frailis}, {Fukazawa}, {Funk}, {Fusco}, {Gargano}, {Gehrels},
  {Germani}, {Giglietto}, {Giordano}, {Glanzman}, {Godfrey}, {Gotthelf},
  {Grenier}, {Grondin}, {Grove}, {Guillemot}, {Guiriec}, {Hanabata}, {Harding},
  {Hays}, {Hobbs}, {Horan}, {Hughes}, {J{\'o}hannesson}, {Johnson}, {Johnson},
  {Johnson}, {Johnston}, {Kamae}, {Kanai}, {Kanbach}, {Katagiri}, {Kataoka},
  {Kawai}, {Keith}, {Kerr}, {Kn{\"o}dlseder}, {Kuss}, {Lande}, {Latronico},
  {Lemoine-Goumard}, {Llena Garde}, {Longo}, {Loparco}, {Lott}, {Lovellette},
  {Lubrano}, {Makeev}, {Manchester}, {Marelli}, {Mazziotta}, {McEnery},
  {Michelson}, {Mitthumsiri}, {Mizuno}, {Moiseev}, {Monte}, {Monzani},
  {Morselli}, {Moskalenko}, {Murgia}, {Nakamori}, {Nolan}, {Norris}, {Nuss},
  {Ohno}, {Ohsugi}, {Omodei}, {Orlando}, {Ormes}, {Paneque}, {Panetta},
  {Parent}, {Pelassa}, {Pepe}, {Pesce-Rollins}, {Piron}, {Porter}, {Rain{\`o}},
  {Rando}, {Razzano}, {Rea}, {Reimer}, {Reimer}, {Reposeur}, {Rodriguez},
  {Romani}, {Roth}, {Ryde}, {Sadrozinski}, {Sander}, {Saz Parkinson},
  {Sgr{\`o}}, {Siskind}, {Smith}, {Smith}, {Spandre}, {Spinelli}, {Starck},
  {Strickman}, {Suson}, {Takahashi}, {Takahashi}, {Tanaka}, {Thayer}, {Thayer},
  {Thompson}, {Thorsett}, {Tibaldo}, {Torres}, {Tosti}, {Tramacere},
  {Uchiyama}, {Usher}, {Vasileiou}, {Venter}, {Vilchez}, {Vitale}, {Waite},
  {Wang}, {Weltevrede}, {Winer}, {Wood}, {Yang}, {Ylinen}, {Ziegler}, {Fermi
  LAT Collaboration}, \& {Pulsar Timing Consortium}}]{AbdoPWN2010}
{Abdo}, A.~A., {Ackermann}, M., {Ajello}, M., {et~al.} 2010, \apj, 714, 927,
  \dodoi{10.1088/0004-637X/714/1/927}

\bibitem[{Abdo {et~al.}(2013)Abdo, Ajello, Allafort, Baldini, Ballet,
  Barbiellini, Baring, Bastieri, Belfiore, Bellazzini, Bhattacharyya, Bissaldi,
  Bloom, Bonamente, Bottacini, Brandt, Bregeon, Brigida, Bruel, Buehler,
  Burgay, Burnett, Busetto, Buson, Caliandro, Cameron, Camilo, Caraveo,
  Casandjian, Cecchi, Çelik, Charles, Chaty, Chaves, Chekhtman, Chen, Chiang,
  Chiaro, Ciprini, Claus, Cognard, Cohen-Tanugi, Cominsky, Conrad, Cutini,
  D'Ammando, de~Angelis, DeCesar, De~Luca, den Hartog, de~Palma, Dermer,
  Desvignes, Digel, Di~Venere, Drell, Drlica-Wagner, Dubois, Dumora, Espinoza,
  Falletti, Favuzzi, Ferrara, Focke, Franckowiak, Freire, Funk, Fusco, Gargano,
  Gasparrini, Germani, Giglietto, Giommi, Giordano, Giroletti, Glanzman,
  Godfrey, Gotthelf, Grenier, Grondin, Grove, Guillemot, Guiriec, Hadasch,
  Hanabata, Harding, Hayashida, Hays, Hessels, Hewitt, Hill, Horan, Hou,
  Hughes, Jackson, Janssen, Jogler, Jóhannesson, Johnson, Johnson, Johnson,
  Johnson, Johnston, Kamae, Kataoka, Keith, Kerr, Knödlseder, Kramer, Kuss,
  Lande, Larsson, Latronico, Lemoine-Goumard, Longo, Loparco, Lovellette,
  Lubrano, Lyne, Manchester, Marelli, Massaro, Mayer, Mazziotta, McEnery,
  McLaughlin, Mehault, Michelson, Mignani, Mitthumsiri, Mizuno, Moiseev,
  Monzani, Morselli, Moskalenko, Murgia, Nakamori, Nemmen, Nuss, Ohno, Ohsugi,
  Orienti, Orlando, Ormes, Paneque, Panetta, Parent, Perkins, Pesce-Rollins,
  Pierbattista, Piron, Pivato, Pletsch, Porter, Possenti, Rainò, Rando,
  Ransom, Ray, Razzano, Rea, Reimer, Reimer, Renault, Reposeur, Ritz, Romani,
  Roth, Rousseau, Roy, Ruan, Sartori, Saz~Parkinson, Scargle, Schulz, Sgrò,
  Shannon, Siskind, Smith, Spandre, Spinelli, Stappers, Strong, Suson,
  Takahashi, Thayer, Thayer, Theureau, Thompson, Thorsett, Tibaldo, Tibolla,
  Tinivella, Torres, Tosti, Troja, Uchiyama, Usher, Vandenbroucke, Vasileiou,
  Venter, Vianello, Vitale, Wang, Weltevrede, Winer, Wolff, Wood, Wood, Wood,
  \& Yang}]{abdo_second_2013}
Abdo, A.~A., Ajello, M., Allafort, A., {et~al.} 2013, The Astrophysical Journal
  Supplement Series, 208, 17, \dodoi{10.1088/0067-0049/208/2/17}

\bibitem[{{Abdollahi} {et~al.}(2020){Abdollahi}, {Acero}, {Ackermann},
  {Ajello}, {Atwood}, {Axelsson}, {Baldini}, {Ballet}, {Barbiellini},
  {Bastieri}, {Becerra Gonzalez}, {Bellazzini}, {Berretta}, {Bissaldi},
  {Blandford}, {Bloom}, {Bonino}, {Bottacini}, {Brandt}, {Bregeon}, {Bruel},
  {Buehler}, {Burnett}, {Buson}, {Cameron}, {Caputo}, {Caraveo}, {Casandjian},
  {Castro}, {Cavazzuti}, {Charles}, {Chaty}, {Chen}, {Cheung}, {Chiaro},
  {Ciprini}, {Cohen-Tanugi}, {Cominsky}, {Coronado-Bl{\'a}zquez}, {Costantin},
  {Cuoco}, {Cutini}, {D'Ammando}, {DeKlotz}, {de la Torre Luque}, {de Palma},
  {Desai}, {Digel}, {Di Lalla}, {Di Mauro}, {Di Venere}, {Dom{\'\i}nguez},
  {Dumora}, {Fana Dirirsa}, {Fegan}, {Ferrara}, {Franckowiak}, {Fukazawa},
  {Funk}, {Fusco}, {Gargano}, {Gasparrini}, {Giglietto}, {Giommi}, {Giordano},
  {Giroletti}, {Glanzman}, {Green}, {Grenier}, {Griffin}, {Grondin}, {Grove},
  {Guiriec}, {Harding}, {Hayashi}, {Hays}, {Hewitt}, {Horan},
  {J{\'o}hannesson}, {Johnson}, {Kamae}, {Kerr}, {Kocevski}, {Kovac'evic'},
  {Kuss}, {Landriu}, {Larsson}, {Latronico}, {Lemoine-Goumard}, {Li},
  {Liodakis}, {Longo}, {Loparco}, {Lott}, {Lovellette}, {Lubrano}, {Madejski},
  {Maldera}, {Malyshev}, {Manfreda}, {Marchesini}, {Marcotulli},
  {Mart{\'\i}-Devesa}, {Martin}, {Massaro}, {Mazziotta}, {McEnery}, {Mereu},
  {Meyer}, {Michelson}, {Mirabal}, {Mizuno}, {Monzani}, {Morselli},
  {Moskalenko}, {Negro}, {Nuss}, {Ojha}, {Omodei}, {Orienti}, {Orlando},
  {Ormes}, {Palatiello}, {Paliya}, {Paneque}, {Pei}, {Pe{\~n}a-Herazo},
  {Perkins}, {Persic}, {Pesce-Rollins}, {Petrosian}, {Petrov}, {Piron}, {Poon},
  {Porter}, {Principe}, {Rain{\`o}}, {Rando}, {Razzano}, {Razzaque}, {Reimer},
  {Reimer}, {Remy}, {Reposeur}, {Romani}, {Saz Parkinson}, {Schinzel},
  {Serini}, {Sgr{\`o}}, {Siskind}, {Smith}, {Spandre}, {Spinelli}, {Strong},
  {Suson}, {Tajima}, {Takahashi}, {Tak}, {Thayer}, {Thompson}, {Tibaldo},
  {Torres}, {Torresi}, {Valverde}, {Van Klaveren}, {van Zyl}, {Wood},
  {Yassine}, \& {Zaharijas}}]{Fermi_Fourth_2019}
{Abdollahi}, S., {Acero}, F., {Ackermann}, M., {et~al.} 2020, \apjs, 247, 33,
  \dodoi{10.3847/1538-4365/ab6bcb}

\bibitem[{{Ackermann} {et~al.}(2011){Ackermann}, {Ajello}, {Baldini}, {Ballet},
  {Barbiellini}, {Bastieri}, {Bechtol}, {Bellazzini}, {Berenji}, {Bloom},
  {Bonamente}, {Borgland}, {Bouvier}, {Bregeon}, {Brez}, {Brigida}, {Bruel},
  {Buehler}, {Buson}, {Caliandro}, {Cameron}, {Camilo}, {Caraveo},
  {Casandjian}, {Cecchi}, {{\c{C}}elik}, {Charles}, {Chekhtman}, {Cheung},
  {Chiang}, {Ciprini}, {Claus}, {Cognard}, {Cohen-Tanugi}, {Conrad}, {Dermer},
  {de Angelis}, {de Luca}, {de Palma}, {Digel}, {Silva}, {Drell}, {Dubois},
  {Dumora}, {Favuzzi}, {Focke}, {Frailis}, {Fukazawa}, {Funk}, {Fusco},
  {Gargano}, {Germani}, {Giglietto}, {Giommi}, {Giordano}, {Giroletti},
  {Glanzman}, {Godfrey}, {Grenier}, {Grondin}, {Grove}, {Guillemot}, {Guiriec},
  {Hadasch}, {Hanabata}, {Harding}, {Hayashi}, {Hays}, {Hobbs}, {Hughes},
  {J{\'o}hannesson}, {Johnson}, {Johnson}, {Johnston}, {Kamae}, {Katagiri},
  {Kataoka}, {Keith}, {Kerr}, {Kn{\"o}dlseder}, {Kramer}, {Kuss}, {Lande},
  {Latronico}, {Lee}, {Lemoine-Goumard}, {Longo}, {Loparco}, {Lovellette},
  {Lubrano}, {Lyne}, {Makeev}, {Marelli}, {Mazziotta}, {McEnery}, {Mehault},
  {Michelson}, {Mizuno}, {Moiseev}, {Monte}, {Monzani}, {Morselli},
  {Moskalenko}, {Murgia}, {Nakamori}, {Naumann-Godo}, {Nolan}, {Noutsos},
  {Nuss}, {Ohsugi}, {Okumura}, {Ormes}, {Paneque}, {Panetta}, {Parent},
  {Pelassa}, {Pepe}, {Pesce-Rollins}, {Piron}, {Porter}, {Rain{\`o}}, {Rando},
  {Ransom}, {Ray}, {Razzano}, {Rea}, {Reimer}, {Reimer}, {Reposeur}, {Ripken},
  {Ritz}, {Romani}, {Sadrozinski}, {Sander}, {Saz Parkinson}, {Sgr{\`o}},
  {Siskind}, {Smith}, {Smith}, {Spandre}, {Spinelli}, {Strickman}, {Suson},
  {Takahashi}, {Takahashi}, {Tanaka}, {Thayer}, {Thayer}, {Theureau},
  {Thompson}, {Thorsett}, {Tibaldo}, {Torres}, {Tosti}, {Tramacere},
  {Uchiyama}, {Uehara}, {Usher}, {Vandenbroucke}, {Van Etten}, {Vasileiou},
  {Vilchez}, {Vitale}, {Waite}, {Wang}, {Weltevrede}, {Winer}, {Wood}, {Yang},
  {Ylinen}, \& {Ziegler}}]{AckermannPWN2011}
{Ackermann}, M., {Ajello}, M., {Baldini}, L., {et~al.} 2011, \apj, 726, 35,
  \dodoi{10.1088/0004-637X/726/1/35}

\bibitem[{{Atwood} {et~al.}(2009){Atwood}, {Abdo}, {Ackermann}, {Althouse},
  {Anderson}, {Axelsson}, {Baldini}, {Ballet}, {Band}, {Barbiellini},
  {Bartelt}, {Bastieri}, {Baughman}, {Bechtol}, {B{\'e}d{\'e}r{\`e}de},
  {Bellardi}, {Bellazzini}, {Berenji}, {Bignami}, {Bisello}, {Bissaldi},
  {Blandford}, {Bloom}, {Bogart}, {Bonamente}, {Bonnell}, {Borgland},
  {Bouvier}, {Bregeon}, {Brez}, {Brigida}, {Bruel}, {Burnett}, {Busetto},
  {Caliandro}, {Cameron}, {Caraveo}, {Carius}, {Carlson}, {Casandjian},
  {Cavazzuti}, {Ceccanti}, {Cecchi}, {Charles}, {Chekhtman}, {Cheung},
  {Chiang}, {Chipaux}, {Cillis}, {Ciprini}, {Claus}, {Cohen-Tanugi},
  {Condamoor}, {Conrad}, {Corbet}, {Corucci}, {Costamante}, {Cutini}, {Davis},
  {Decotigny}, {DeKlotz}, {Dermer}, {de Angelis}, {Digel}, {do Couto e Silva},
  {Drell}, {Dubois}, {Dumora}, {Edmonds}, {Fabiani}, {Farnier}, {Favuzzi},
  {Flath}, {Fleury}, {Focke}, {Funk}, {Fusco}, {Gargano}, {Gasparrini},
  {Gehrels}, {Gentit}, {Germani}, {Giebels}, {Giglietto}, {Giommi}, {Giordano},
  {Glanzman}, {Godfrey}, {Grenier}, {Grondin}, {Grove}, {Guillemot}, {Guiriec},
  {Haller}, {Harding}, {Hart}, {Hays}, {Healey}, {Hirayama}, {Hjalmarsdotter},
  {Horn}, {Hughes}, {J{\'o}hannesson}, {Johansson}, {Johnson}, {Johnson},
  {Johnson}, {Johnson}, {Kamae}, {Katagiri}, {Kataoka}, {Kavelaars}, {Kawai},
  {Kelly}, {Kerr}, {Klamra}, {Kn{\"o}dlseder}, {Kocian}, {Komin}, {Kuehn},
  {Kuss}, {Landriu}, {Latronico}, {Lee}, {Lee}, {Lemoine-Goumard}, {Lionetto},
  {Longo}, {Loparco}, {Lott}, {Lovellette}, {Lubrano}, {Madejski}, {Makeev},
  {Marangelli}, {Massai}, {Mazziotta}, {McEnery}, {Menon}, {Meurer},
  {Michelson}, {Minuti}, {Mirizzi}, {Mitthumsiri}, {Mizuno}, {Moiseev},
  {Monte}, {Monzani}, {Moretti}, {Morselli}, {Moskalenko}, {Murgia},
  {Nakamori}, {Nishino}, {Nolan}, {Norris}, {Nuss}, {Ohno}, {Ohsugi}, {Omodei},
  {Orlando}, {Ormes}, {Paccagnella}, {Paneque}, {Panetta}, {Parent}, {Pearce},
  {Pepe}, {Perazzo}, {Pesce-Rollins}, {Picozza}, {Pieri}, {Pinchera}, {Piron},
  {Porter}, {Poupard}, {Rain{\`o}}, {Rando}, {Rapposelli}, {Razzano}, {Reimer},
  {Reimer}, {Reposeur}, {Reyes}, {Ritz}, {Rochester}, {Rodriguez}, {Romani},
  {Roth}, {Russell}, {Ryde}, {Sabatini}, {Sadrozinski}, {Sanchez}, {Sander},
  {Sapozhnikov}, {Parkinson}, {Scargle}, {Schalk}, {Scolieri}, {Sgr{\`o}},
  {Share}, {Shaw}, {Shimokawabe}, {Shrader}, {Sierpowska-Bartosik}, {Siskind},
  {Smith}, {Smith}, {Spandre}, {Spinelli}, {Starck}, {Stephens}, {Strickman},
  {Strong}, {Suson}, {Tajima}, {Takahashi}, {Takahashi}, {Tanaka}, {Tenze},
  {Tether}, {Thayer}, {Thayer}, {Thompson}, {Tibaldo}, {Tibolla}, {Torres},
  {Tosti}, {Tramacere}, {Turri}, {Usher}, {Vilchez}, {Vitale}, {Wang},
  {Watters}, {Winer}, {Wood}, {Ylinen}, \& {Ziegler}}]{Atwood_Mission_2009}
{Atwood}, W.~B., {Abdo}, A.~A., {Ackermann}, M., {et~al.} 2009, \apj, 697,
  1071, \dodoi{10.1088/0004-637X/697/2/1071}

\bibitem[{{Auchettl} {et~al.}(2014){Auchettl}, {Slane}, \&
  {Castro}}]{Auchettl2014}
{Auchettl}, K., {Slane}, P., \& {Castro}, D. 2014, \apj, 783, 32,
  \dodoi{10.1088/0004-637X/783/1/32}

\bibitem[{Bochow(2011)}]{Bochow2011ASS}
Bochow, A. 2011, PhD thesis

\bibitem[{{Caswell} {et~al.}(1975){Caswell}, {Clark}, \&
  {Crawford}}]{Caswell1975}
{Caswell}, J.~L., {Clark}, D.~H., \& {Crawford}, D.~F. 1975, Australian Journal
  of Physics Astrophysical Supplement, 37, 39

\bibitem[{{Cordes} \& {Lazio}(2003)}]{Cordes2003}
{Cordes}, J.~M., \& {Lazio}, T.~J.~W. 2003, arXiv e-prints, astro.
\newblock \doarXiv{astro-ph/0301598}

\bibitem[{{Cui} {et~al.}(2018){Cui}, {Yeung}, {Tam}, \&
  {P{\"u}hlhofer}}]{Cui2018}
{Cui}, Y., {Yeung}, P. K.~H., {Tam}, P.~H.~T., \& {P{\"u}hlhofer}, G. 2018,
  \apj, 860, 69, \dodoi{10.3847/1538-4357/aac37b}

\bibitem[{{Ginzburg} \& {Syrovatskii}(1964)}]{Ginzburg1964}
{Ginzburg}, V.~L., \& {Syrovatskii}, S.~I. 1964, {The Origin of Cosmic Rays}

\bibitem[{{Halpern} \& {Gotthelf}(2010{\natexlab{a}})}]{Halpern2010b}
{Halpern}, J.~P., \& {Gotthelf}, E.~V. 2010{\natexlab{a}}, \apj, 725, 1384,
  \dodoi{10.1088/0004-637X/725/1/1384}

\bibitem[{{Halpern} \& {Gotthelf}(2010{\natexlab{b}})}]{Halpern2010a}
---. 2010{\natexlab{b}}, \apj, 709, 436, \dodoi{10.1088/0004-637X/709/1/436}

\bibitem[{{Hobbs} {et~al.}(2004){Hobbs}, {Lyne}, {Kramer}, {Martin}, \&
  {Jordan}}]{Hobbs2004}
{Hobbs}, G., {Lyne}, A.~G., {Kramer}, M., {Martin}, C.~E., \& {Jordan}, C.
  2004, \mnras, 353, 1311, \dodoi{10.1111/j.1365-2966.2004.08157.x}

\bibitem[{{Jackson} {et~al.}(2006){Jackson}, {Rathborne}, {Shah}, {Simon},
  {Bania}, {Clemens}, {Chambers}, {Johnson}, {Dormody}, {Lavoie}, \&
  {Heyer}}]{Jackson2006}
{Jackson}, J.~M., {Rathborne}, J.~M., {Shah}, R.~Y., {et~al.} 2006, \apjs, 163,
  145, \dodoi{10.1086/500091}

\bibitem[{{Kafexhiu} {et~al.}(2014){Kafexhiu}, {Aharonian}, {Taylor}, \&
  {Vila}}]{Kafexhiu2014}
{Kafexhiu}, E., {Aharonian}, F., {Taylor}, A.~M., \& {Vila}, G.~S. 2014, \prd,
  90, 123014, \dodoi{10.1103/PhysRevD.90.123014}

\bibitem[{{Kelner} {et~al.}(2006){Kelner}, {Aharonian}, \&
  {Bugayov}}]{Kelner2006}
{Kelner}, S.~R., {Aharonian}, F.~A., \& {Bugayov}, V.~V. 2006, \prd, 74,
  034018, \dodoi{10.1103/PhysRevD.74.034018}

\bibitem[{{Kilpatrick} {et~al.}(2016){Kilpatrick}, {Bieging}, \&
  {Rieke}}]{Kilpatrick2016}
{Kilpatrick}, C.~D., {Bieging}, J.~H., \& {Rieke}, G.~H. 2016, \apj, 816, 1,
  \dodoi{10.3847/0004-637X/816/1/1}

\bibitem[{{Li} {et~al.}(2018){Li}, {Torres}, {Lin}, {Grondin}, {Kerr},
  {Lemoine-Goumard}, \& {de O{\~n}a Wilhelmi}}]{LiPWN2018}
{Li}, J., {Torres}, D.~F., {Lin}, T.~T., {et~al.} 2018, \apj, 858, 84,
  \dodoi{10.3847/1538-4357/aabac9}

\bibitem[{{Panaitescu}(2017)}]{Panaitescu2017}
{Panaitescu}, A. 2017, \apj, 837, 13, \dodoi{10.3847/1538-4357/837/1/13}

\bibitem[{{Pellizzoni} {et~al.}(2002){Pellizzoni}, {Mereghetti}, {Tavani},
  {Chen}, {Giuliani}, \& {Vercellone}}]{Pellizzoni2002}
{Pellizzoni}, A., {Mereghetti}, S., {Tavani}, M., {et~al.} 2002, in 34th COSPAR
  Scientific Assembly, Vol.~34, 2660.
\newblock \doarXiv{astro-ph/0212202}

\bibitem[{{Ptuskin} \& {Zirakashvili}(2003)}]{Ptuskin2003}
{Ptuskin}, V.~S., \& {Zirakashvili}, V.~N. 2003, \aap, 403, 1,
  \dodoi{10.1051/0004-6361:20030323}

\bibitem[{{Ptuskin} \& {Zirakashvili}(2005)}]{Ptuskin2005}
---. 2005, \aap, 429, 755, \dodoi{10.1051/0004-6361:20041517}

\bibitem[{{Rea} {et~al.}(2014){Rea}, {Vigan{\`o}}, {Israel}, {Pons}, \&
  {Torres}}]{Rea2014}
{Rea}, N., {Vigan{\`o}}, D., {Israel}, G.~L., {Pons}, J.~A., \& {Torres}, D.~F.
  2014, \apjl, 781, L17, \dodoi{10.1088/2041-8205/781/1/L17}

\bibitem[{{Rho} \& {Petre}(1998)}]{Rho1998}
{Rho}, J., \& {Petre}, R. 1998, \apjl, 503, L167, \dodoi{10.1086/311538}

\bibitem[{Roberts(2004)}]{roberts2004pwn}
Roberts, M.~S.~E. 2004, McGill University, Montreal, Quebec, Canada (available
  on the world-wide-web at http://www. physics. mcgill. ca/\~{} pulsar/pwncat.
  html)

\bibitem[{{Sato} {et~al.}(2016){Sato}, {Koyama}, {Lee}, \&
  {Takahashi}}]{Sato2016}
{Sato}, T., {Koyama}, K., {Lee}, S.-H., \& {Takahashi}, T. 2016, \pasj, 68, S8,
  \dodoi{10.1093/pasj/psv131}

\bibitem[{{Seaquist} \& {Gilmore}(1982)}]{SeaquistGilmore1982}
{Seaquist}, E.~R., \& {Gilmore}, W.~S. 1982, \aj, 87, 378,
  \dodoi{10.1086/113109}

\bibitem[{{Seward} {et~al.}(2003){Seward}, {Slane}, {Smith}, \&
  {Sun}}]{Seward2003}
{Seward}, F.~D., {Slane}, P.~O., {Smith}, R.~K., \& {Sun}, M. 2003, \apj, 584,
  414, \dodoi{10.1086/345600}

\bibitem[{{Seward} \& {Velusamy}(1995)}]{SewardVelusamy1995}
{Seward}, F.~D., \& {Velusamy}, T. 1995, \apj, 439, 715, \dodoi{10.1086/175211}

\bibitem[{{Stanimirovi{\'c}} {et~al.}(2003){Stanimirovi{\'c}}, {Weisberg},
  {Dickey}, {de la Fuente}, {Devine}, {Hedden}, \&
  {Anderson}}]{Stanimirovic2003}
{Stanimirovi{\'c}}, S., {Weisberg}, J.~M., {Dickey}, J.~M., {et~al.} 2003,
  \apj, 592, 953, \dodoi{10.1086/375779}

\bibitem[{{Sun} {et~al.}(2004){Sun}, {Seward}, {Smith}, \& {Slane}}]{Sun2004}
{Sun}, M., {Seward}, F.~D., {Smith}, R.~K., \& {Slane}, P.~O. 2004, \apj, 605,
  742, \dodoi{10.1086/382666}

\bibitem[{{Suzuki} {et~al.}(2020{\natexlab{a}}){Suzuki}, {Bamba}, {Enokiya},
  {Yamaguchi}, {Plucinsky}, \& {Odaka}}]{Suzuki2020a}
{Suzuki}, H., {Bamba}, A., {Enokiya}, R., {et~al.} 2020{\natexlab{a}}, \apj,
  893, 147, \dodoi{10.3847/1538-4357/ab80ba}

\bibitem[{{Suzuki} {et~al.}(2018){Suzuki}, {Bamba}, {Nakazawa}, {Furuta},
  {Sawada}, {Yamazaki}, \& {Koyama}}]{Suzuki2018}
{Suzuki}, H., {Bamba}, A., {Nakazawa}, K., {et~al.} 2018, \pasj, 70, 75,
  \dodoi{10.1093/pasj/psy069}

\bibitem[{{Suzuki} {et~al.}(2020{\natexlab{b}}){Suzuki}, {Bamba}, {Yamazaki},
  \& {Ohira}}]{Suzuki2020b}
{Suzuki}, H., {Bamba}, A., {Yamazaki}, R., \& {Ohira}, Y. 2020{\natexlab{b}},
  \pasj, 72, 72, \dodoi{10.1093/pasj/psaa061}

\bibitem[{{Thoudam}(2007)}]{Thoudam2007}
{Thoudam}, S. 2007, \mnras, 378, 48, \dodoi{10.1111/j.1365-2966.2007.11705.x}

\bibitem[{{Tibolla} {et~al.}(2011){Tibolla}, {Mannheim}, {Els{\"a}sser}, \&
  {Kaufmann}}]{Tibolla2011}
{Tibolla}, O., {Mannheim}, K., {Els{\"a}sser}, D., \& {Kaufmann}, S. 2011,
  arXiv e-prints, arXiv:1111.1634.
\newblock \doarXiv{1111.1634}

\bibitem[{{Velusamy} {et~al.}(1991){Velusamy}, {Becker}, \&
  {Seward}}]{Velusamy1991}
{Velusamy}, T., {Becker}, R.~H., \& {Seward}, F.~D. 1991, \aj, 102, 676,
  \dodoi{10.1086/115901}

\bibitem[{Zhang(2003)}]{Zhang2003}
Zhang, B. 2003, in {International Workshop on Strong Magnetic Fields and
  Neutron Star}, 83

\bibitem[{{Zhou} {et~al.}(2014){Zhou}, {Chen}, {Li}, {Safi-Harb}, {Mendez},
  {Terada}, {Sun}, \& {Ge}}]{Zhou2014}
{Zhou}, P., {Chen}, Y., {Li}, X.-D., {et~al.} 2014, \apjl, 781, L16,
  \dodoi{10.1088/2041-8205/781/1/L16}

\bibitem[{{Zhou} {et~al.}(2016){Zhou}, {Chen}, {Safi-Harb}, {Zhou}, {Sun},
  {Zhang}, \& {Zhang}}]{Zhou2016}
{Zhou}, P., {Chen}, Y., {Safi-Harb}, S., {et~al.} 2016, \apj, 831, 192,
  \dodoi{10.3847/0004-637X/831/2/192}

\bibitem[{{Zirakashvili} \& {Ptuskin}(2008)}]{Zirakashvili2008}
{Zirakashvili}, V.~N., \& {Ptuskin}, V.~S. 2008, \apj, 678, 939,
  \dodoi{10.1086/529580}

\bibitem[{{Zubrin} \& {Shulga}(2008)}]{Zubrin2008}
{Zubrin}, S.~Y., \& {Shulga}, V.~M. 2008, in Young Scientists 15th Proceedings,
  ed. V.~Y. {Choliy} \& G.~{Ivashchenko}, 41--43

\end{thebibliography}



\end{document}